\documentclass[11pt]{article}

\usepackage{amssymb,amsthm,amsmath}
\usepackage{mathrsfs}
\usepackage{graphicx}
\usepackage[FIGTOPCAP,nooneline]{subfigure}
\usepackage[top=35mm,bottom=35mm]{geometry}
\usepackage{color}

\makeatletter
    
    \@addtoreset{equation}{section}
\makeatother

\theoremstyle{definition}
\newtheorem{theorem}{Theorem}[section]

\newtheorem{remark}[theorem]{Remark}

\newcommand{\eps}{\varepsilon}
\newcommand{\R}{{\mathbb R}}
\newcommand{\bs}[1]{\boldsymbol{#1}}

\usepackage{authblk}

\title{Enstrophy variations in the collapsing process of point vortices}

\author{Takeshi Gotoda \footnote{Department of Mathematical and Computing Science, Institute of Science Tokyo, 2-12-1 Ookayama, Meguro-ku, Tokyo, JAPAN  E-mail: gotoda.t.dde7@m.isct.ac.jp} }

\date{}

\begin{document}


\maketitle

\begin{abstract}

We investigate enstrophy variations by collapse of point vortices in an inviscid flow and, in particular, focus on the enstrophy dissipation that is a significant property characterizing 2D turbulent flows. Point vortex is an ideal vortex whose vorticity is concentrated on a point and the dynamics of point vortices on an inviscid flow is described by the point-vortex system. The point-vortex system has self-similar collapsing solutions, which are expected to cause the anomalous enstrophy dissipation, but this collapsing process of point vortices cannot be described by the 2D Euler equations.
In this study, we consider point-vortex solutions of the 2D filtered Euler equations, which are a regularized model of the 2D Euler equations, and the filtered-point-vortex system describing the dynamics of them. The preceding studies \cite{GS2,GS3} have proven that there exist solutions to the three filtered-point-vortex system such that they converge to self-similar collapsing orbits of the point-vortex system and dissipate the enstrophy at the event of collapse in the zero limit of a filter scale. In this study, we numerically show that the enstrophy dissipation by the collapse of point vortices could occur for the four and five vortex problems.

\end{abstract}

\section{Introduction}

In 2D turbulent flows at high Reynolds number, there appears an inconsistency in flow regularity between inviscid limits of viscous flows and non-viscous ones: the dissipation of the enstrophy, which is the $L^2$ norm of the scalar vorticity, in the inviscid limit gives rise to the inertial range of the energy density spectrum corresponding to the forward enstrophy cascade in 2D turbulence \cite{Batchelor,Kraichnan,Leith}, but smooth solutions to the 2D incompressible Euler equations conserve the enstrophy. This inconsistency insists that turbulent flows subject to the 2D Navier-Stokes equations converge to non-smooth flows governed by the 2D Euler equations in the inviscid limit. Our motivation of the present study is to investigate non-smooth solutions of inviscid models and understand the physical mechanism of 2D turbulent flows.

To construct non-smooth solutions dissipating the enstrophy to the 2D Euler equations, we have to deal with less regular vorticity that we call singular vorticity and, according to the preceding study \cite{Eyink}, the enstrophy dissipation could occur for  the vorticity such as distributions in the space of finite Radon measures. However, the global well-posedness of the 2D Euler equations has not been established for vorticity described by finite Radon measures. To overcome this difficulty, we regularize the Euler equations on the basis of a spatial filtering. We call the regularized model \textit{the filtered-Euler equations}, which are a generalized model of the Euler-$\alpha$ equations and the vortex blob regularization. The advantage of considering the filtered-model is the existence of a unique global weak solution for initial vorticity in the space of finite Radon measures \cite{G1}. In addition, the 2D filtered-Euler equations converges to the 2D Euler equations in the zero limit of the filter parameter for certain classes of initial vorticity \cite{G1,G2}. Our strategy for constructing enstrophy dissipating solutions is to find a unique global weak solution to the 2D filtered-Euler equations that dissipates the enstrophy in the zero limit of the filter parameter.

Another aim of this study is making it clear what kind of vortex motions causes the enstrophy dissipation. For this purpose, we consider the vorticity represented by a $\delta$-measure, which we call \textit{point vortex}, as initial vorticity in the space of finite Radon measures, since the dynamics of point vortices is described by their orbits and it is enough to trace them mathematically or numerically.
Although the existence of a weak solution has not been established for the 2D Euler equations with point-vortex initial vorticity, {\it the point-vortex system} is known as a model describing the dynamics of point vortices on the 2D Euler flow formally. One of the notable features of the point-vortex system is the existence of self-similar collapsing solutions, that is, point vortices simultaneously collide with each other at a finite time. The mechanism of collapse of multiple vortices plays an important role to understand fluid phenomena. For example, the dynamics of point vortices is utilized as a simple model of the 2D turbulence and collapse of point vortices is considered as an elementary process in the 2D turbulence kinetics \cite{Benzi,Carnevale,Leoncini,Novikov(a),Siggia,Weiss}. On the other hand, the 2D filtered-Euler equations have a global weak solution for point-vortex initial vorticity and the evolution of their point-vortex solution is described by ordinary differential equations called {\it the filtered-point-vortex system}. Our aim is to show the existence of a solution to the filtered-point-vortex system that causes the anomalous enstrophy dissipation via collapse of point vortices in the zero limit of the filter parameter.

As the first attempt of constructing a enstrophy dissipating solution of point vortices, the three vortex problem in the 2D Euler-$\alpha$ equations has been considered in \cite{Sakajo}. The dynamics of point-vortex solutions to the 2D Euler-$\alpha$ equations is described by the $\alpha$-point-vortex system. The author in \cite{Sakajo} has considered the three $\alpha$-point-vortex system with the initial data leading to a self-similar collapse in the three point-vortex system, and shown with a help of numerical computations that the solution to the three $\alpha$-point-vortex system converges to a self-similar collapsing orbit in the zero limit of the filter parameter and a variational part of the total enstrophy dissipates at the event of triple collapse.
This result has been made mathematically rigorous and it has been shown that the enstrophy dissipation by the triple collapse occurs for a wider set of initial data in \cite{GS2}. Then, the results in \cite{GS2, Sakajo} has been mathematically extended to the general 2D filtered-Euler equations in \cite{GS3}.
In this paper, we consider the four and five vortex problem in the filtered-point-vortex system. Although we have not found explicit formulae for self-similar collapsing solutions to the four and five point-vortex system in general, an example of initial data leading to self-similar collapse has been found in \cite{Novikov}. Thus, we numerically show that solutions to the four and five filtered-point-vortex system with initial data in \cite{Novikov} lead to self-similar collapse and the enstrophy dissipation in the zero limit of the filter parameter.

This paper is organized as follows. In Section~\ref{sec:PVS}, we briefly review the point-vortex system. In order to compare with the filtered-point-vortex system, we derive the point-vortex system from the 2D Euler equation on the basis of the Lagrangian flow map. Then, we introduce the definition of self-similar motions and examples of exact self-similar collapsing solutions to the point-vortex system.
In Section~\ref{sec:FPVS}, we derive the filtered-point-vortex system from the 2D filtered-Euler equations and see fundamental properties. After introducing the variations of the enstrophy and the energy to the filtered point-vortex system, we mention preceding results about the enstrophy dissipation via self-similar collapse of three point vortices.
The main results are shown in Section~\ref{sec:main}. In this section, we first see numerical solutions to the three $\alpha$-point-vortex system. Although the three vortex problem has been already settled with a mathematical rigor, we see the detailed process of the enstrophy dissipation in comparison with the mathematical theory. Then, we numerically show that self-similar collapse of four and five point vortices causes the enstrophy dissipation by considering the zero limit of the filter scale.
Section~\ref{sec:concluding} is devoted to concluding remarks.

\section{The point vortex system}
\label{sec:PVS}

\subsection{Derivation based on the 2D Euler equations}
\label{subsec:formulation}
In this section, we review the formulation of the point-vortex system and its fundamental properties. We start by considering the 2D Euler equations as an inviscid model:
\begin{equation}
\partial_t \boldsymbol{v} + (\boldsymbol{v} \cdot \nabla) \boldsymbol{v} + \nabla p = 0, \qquad  \nabla \cdot \boldsymbol{v} = 0, \label{eq:E}
\end{equation}
where unknown functions $\boldsymbol{v}(\boldsymbol{x},t) = (v_1(\boldsymbol{x},t),v_2(\boldsymbol{x},t))$ and $p(\boldsymbol{x},t)$ describe a velocity field and a pressure function, respectively. Taking the $\operatorname{curl}$ of \eqref{eq:E}, we obtain a transport equation for the vorticity $q := \operatorname{curl} \boldsymbol{v} = \partial_{x_1} v_2 - \partial_{x_2} v_1$:
\begin{equation}
\partial_t q + (\boldsymbol{v} \cdot \nabla) q = 0, \label{eq:VE}
\end{equation}
which we call the vorticity form of \eqref{eq:E}, and the Biot-Savart law gives
\begin{equation*}
\boldsymbol{v}(\bs{x},t) = (\boldsymbol{K} \ast q)(\bs{x},t) := \int_{\R^2} \bs{K}(\bs{x}- \bs{y}) q(\bs{y},t) d \bs{y}, \qquad \boldsymbol{K}(\boldsymbol{x}) := \frac{1}{2 \pi}\frac{\boldsymbol{x}^\perp}{|\boldsymbol{x}|^2},
\end{equation*}
where $\bs{x}^\perp := (- x_2, x_1)$. The initial value problem of \eqref{eq:VE} in $\R^2$ has a unique global weak solution for initial vorticity $q_0 \in L^1(\R^2) \cap L^\infty(\R^2)$ \cite{Yudovich}. Then, owing to the uniqueness, we have the Lagrangian flow map $\boldsymbol{\eta}$ governed by
\begin{equation}
\partial_t \boldsymbol{\eta}(\boldsymbol{x}, t) = \boldsymbol{v} \left( \boldsymbol{\eta}(\boldsymbol{x}, t), t \right), \qquad \boldsymbol{\eta}(\boldsymbol{x}, 0) = \boldsymbol{x}. \label{eq:FME}
\end{equation}
Note that a unique solution of \eqref{eq:FME} yields a solution of \eqref{eq:VE} via $q(\boldsymbol{x},t) = q_0(\boldsymbol{\eta}(\boldsymbol{x}, -t))$.
The existence of a global weak solution to \eqref{eq:VE} has been extended to the case of $q_0 \in L^1(\R^2) \cap L^p(\R^2)$ with $p > 1$ \cite{DiPerna} and less regular vorticity $q_0 \in \mathcal{M}(\mathbb{R}^2) \cap H_{\mathrm{loc}}^{-1}(\mathbb{R}^2)$ \cite{Delort, Majda},
where $\mathcal{M}(\mathbb{R}^2)$ and $H_{\mathrm{loc}}^{-1}(\mathbb{R}^2)$ denote the space of finite Radon measures and the Sobolev space, respectively.

In this paper, we focus on point-vortex initial vorticity, which is represented by
\begin{equation}
q_0(\boldsymbol{x}) = \sum_{m=1}^N \Gamma_m \ \delta(\boldsymbol{x} - \boldsymbol{k}_m), \label{PVI}
\end{equation}
where $N \in \mathbb{N}$ is the number of point vortices and $\delta(\bs{x})$ is the Dirac delta function. The given constants $\Gamma_m \in \R \setminus \{0 \}$ and $\bs{k}_m \in \R^2$ denote the strength and the initial position of $m$-th point vortex, respectively. Unfortunately, the solvability of \eqref{eq:VE} has not been established for initial vorticity \eqref{PVI} in general. In what follows, we formally derive the governing equations of point vortices from \eqref{eq:VE} by assuming that point vortices are convected by the flow map \eqref{eq:FME} and the solution of \eqref{eq:VE} is described by
\begin{equation*}
q(\boldsymbol{x},t) = \sum_{m=1}^N \Gamma_m \ \delta(\boldsymbol{x} - \bs{\eta}(\bs{k}_m,t)).
\end{equation*}
For simplicity, we set $\bs{x}_m(t) := \bs{\eta}(\bs{k}_m,t)$. Then, \eqref{eq:FME} and the Biot-Savart law yield
\begin{equation*}
\frac{\mbox{d}}{\mbox{d}t} \boldsymbol{x}_m(t) = \sum_{n=1}^N \Gamma_n \int \boldsymbol{K} ( \boldsymbol{x}_m(t) - \boldsymbol{y})\ \delta(\boldsymbol{y} - \boldsymbol{x}_n(t)) d\boldsymbol{y} \sim \sum_{n\neq m}^N \Gamma_n \boldsymbol{K} (\boldsymbol{x}_m(t) - \boldsymbol{x}_n(t) ).
\end{equation*}
Note that the last approximation is not a mathematically rigorous calculation but a formal one, since the kernel $\bs{K}$ is not bounded at the origin. Introducing complex positions of point vortices $z_m(t) := x_m(t) + i y_m(t)$ for convenience of notation, we find {\it the point-vortex (PV) system}:
\begin{equation}
\dfrac{\mbox{d} z_m}{\mbox{d}t} = \frac{-1}{2 \pi i} \sum_{n\neq m}^N \dfrac{\Gamma_n}{\overline{z}_m - \overline{z}_n}, \qquad z_m(0) = k_m \label{eq:PV}
\end{equation}
for $m = 1,\cdots, N$, where $\overline{z}_m$ denotes the complex conjugate of $z_m$ and $k_m$ does the complex position of $\boldsymbol{k}_m \in \mathbb{R}^2$. The PV system is formulated as a Hamiltonian system \cite{Kirchhoff}: motions of point vortices $\{(x_m(t), y_m(t))\}_{m=1}^N$ are described by
\begin{equation*}
\Gamma_m\dfrac{\mbox{d} x_m}{\mbox{d} t} = \dfrac{\partial \mathscr{H}}{\partial y_m}, \qquad \Gamma_m\dfrac{\mbox{d} y_m}{\mbox{d} t} = -\dfrac{\partial \mathscr{H}}{\partial x_m} \label{pv-h}
\end{equation*}
with the Hamiltonian,
\begin{equation*}
\mathscr{H} := - \frac{1}{2 \pi} \sum_{m=1}^N \sum_{n=m+1}^N \Gamma_m \Gamma_n \log{l_{mn}},
\end{equation*}
where $l_{mn}(t) := |z_m(t) - z_n(t)|$. In addition to the Hamiltonian $\mathscr{H}$, the PV system has the following invariant quantities:
\begin{align*}
P + i Q := \sum_{m=1}^N \Gamma_m x_m + i \sum_{m=1}^N \Gamma_m y_m, \qquad I := \sum_{m=1}^N \Gamma_m |z_m|^2,
\end{align*}
and these quantities yield another invariant,
\begin{equation*}
M := \sum_{m=1}^N \sum_{n=m+1}^N \Gamma_m \Gamma_n l_{mn}^2 = 2(\Gamma I - P^2 - Q^2),
\end{equation*}
where $\Gamma := \sum_{m=1}^N \Gamma_m$.
Considering these invariants, we find that the PV system \eqref{eq:PV} with $N \leq 3$ is integrable for any $\Gamma_m \in \mathbb{R}\setminus\{0\}$ and the system with $N = 4$ is integrable when $\Gamma = 0$ holds, see \cite{Newton} for details. The system is no longer integrable for $N = 4$ with $\Gamma \neq 0$ and $N \geq 5$: the dynamics of point vortices could be chaotic.
It is useful to consider the evolution of mutual distances $l_{mn}$ for $N \geq 3$, which is governed by
\begin{equation*}
\frac{\mbox{d}}{\mbox{d} t} l_{mn}^2 = \frac{2}{\pi} \sum_{l \neq n \neq m}^N \Gamma_l \sigma_{mnl} A_{mnl} \left( \frac{1}{l_{nl}^2} - \frac{1}{l_{lm}^2} \right), \qquad m \neq n.
\end{equation*}
Here, $A_{mnl}$ denotes the area of the triangle formed by $(z_m, z_n, z_l)$, which is expressed by
\begin{equation*}
A_{mnl} = \frac{1}{4} \left[ 2 \left( l_{mn}^2 l_{nl}^2 + l_{nl}^2 l_{lm}^2 + l_{lm}^2 l_{mn}^2 \right) - l_{mn}^4 - l_{nl}^4 - l_{lm}^4 \right]^{\frac{1}{2}},
\end{equation*}
and $\sigma_{mnl}$ is the sign of the area, that is, $\sigma_{mnl} = 1$ if the indices $(m, n, l)$ at the vertices of the triangle appear counterclockwise and $\sigma_{mnl} = -1$ if they do clockwise.

\subsection{Self-similar solutions}
\label{subsec:self-similar}

Self-similar motions of point vortices are described by the form of
\begin{equation}
z_m(t) = k_m f(t), \qquad f(t) := r(t) e^{i \theta(t)}, \label{hyp-self-similar}
\end{equation}
where $r \geq 0$ and $\theta \in \mathbb{R}$ satisfy $r(0) = 1$ and $\theta(0) = 0$ \cite{Kimura}. Here, owing to the translational invariance of the PV system, we fix the center of point vortices to the origin, that is, $P=Q=0$.
Substituting \eqref{hyp-self-similar} into \eqref{eq:PV}, we find
\begin{equation*}
2 \pi \dfrac{\mbox{d}f}{\mbox{d}t} \overline{f} = \frac{i}{k_m} \sum_{n\neq m}^N  \dfrac{\Gamma_n}{\overline{k}_m - \overline{k}_n}.
\end{equation*}
Thus, the existence of a self-similar solution is equivalent to the existence of an initial configuration $\{ k_m \}_{m=1}^N$ for which there exist constants $A$, $B \in \mathbb{R}$, which are independent of $m$, such that they satisfy
\begin{equation}
A + i B = \frac{i}{2 \pi k_m} \sum_{n\neq m}^N  \dfrac{\Gamma_n}{\overline{k}_m - \overline{k}_n} \label{def-AB}
\end{equation}
for any $m = 1,\cdots,N$. For self-similar solutions, the constants $A$ and $B$ are expressed by
\begin{align*}
A &= \frac{1}{\pi l_{mn}^2(0)} \sum_{l \neq n \neq m}^N \Gamma_l \sigma_{mnl} A_{mnl} \left( \frac{1}{l_{nl}^2(0)} - \frac{1}{l_{lm}^2(0)} \right), \\
B &= \frac{1}{2\pi l_{mn}^2(0)} \left[ \Gamma_m + \Gamma_n + (\boldsymbol{k}_m - \boldsymbol{k}_n) \cdot \sum_{l \neq n \neq m}^N \Gamma_l \left( \frac{\boldsymbol{k}_l - \boldsymbol{k}_n}{l_{nl}^2(0)} - \frac{\boldsymbol{k}_l - \boldsymbol{k}_m}{l_{lm}^2(0)} \right) \right]
\end{align*}
for any $m \neq n$, where $\boldsymbol{k}_m := (Re[k_m],Im[k_m])$, see \cite{G3}. For the case of $A \neq 0$, the self-similar solution of the PV system is explicitly described by
\begin{equation*}
z_m(t) = k_m \sqrt{2At + 1} \exp{\left[ i \dfrac{B}{2A}\log{(2At + 1)} \right]}, \quad m = 1,\cdots,N,
\end{equation*}
and the mutual distances are given by $l_{mn}(t) = l_{mn}(0) \sqrt{2At + 1}$
for $m \neq n$ \cite{Kimura}. Thus, the point vortices simultaneously collide at the origin with the time
\begin{equation}
t_c := - \frac{1}{2A}, \label{def:tc}
\end{equation}
which is called {\it the collapse time}. We also call self-similar motions with $A < 0$ {\it collapsing} and those with $A > 0$ {\it expanding} in the positive direction of time. For the initial position satisfying $A =0$, the corresponding self-similar solution is a relative equilibrium in the form $z_m(t) = k_m e^{i B t}$ that rotates rigidly about their center of vorticity with the angular velocity $B$. It is easily confirmed that self-similar solutions with $A \neq 0$ satisfy $I = M = 0$ and
\begin{equation*}
\Gamma_H := \sum_{m=1}^N \sum_{n=m+1}^N \Gamma_m \Gamma_n = 0,
\end{equation*}
which follows from the invariance of the Hamiltonian. Note that we may fix the $N$-th point vortex of the initial configuration to $z_N = 1$ on the real axis, since the PV system has rotational and scaling invariance.

\subsection{Exact solutions for self-similar collapse}
\label{subsec:exact-solutions}

In the three PV system, it is known that $\Gamma_H = M = 0$ is a necessary and sufficient condition for the self-similar collapse \cite{Aref,Kimura,Newton}, and any partial collapse does not occur. Note that $\Gamma_H = 0$ allows us to assume that
$\Gamma_1$, $\Gamma_2$ have a same sign and $\Gamma_3$ does the opposite one without loss of generality: $\Gamma_3$ is replaced by $ - \Gamma_1 \Gamma_2 /(\Gamma_1 + \Gamma_2)$. Under the condition $\Gamma_H = M = 0$, initial configurations of self-similar solutions are expressed by
\begin{equation}
k_1 = \dfrac{\Gamma_1 \Gamma_2}{(\Gamma_1 + \Gamma_2)^2} \left( 1 + \dfrac{\sqrt{\mathcal{R}}}{\Gamma_1} e^{- i \theta} \right), \quad k_2 = \dfrac{\Gamma_1 \Gamma_2}{(\Gamma_1 + \Gamma_2)^2} \left( 1 - \dfrac{\sqrt{\mathcal{R}}}{\Gamma_2} e^{- i \theta} \right), \quad k_3 = 1 \label{init:three}
\end{equation}
for $\theta \in [0, 2\pi)$, where $\mathcal{R} := \Gamma_1^2 + \Gamma_1 \Gamma_2 + \Gamma_2^2$, see \cite{Kimura}. Then, their mutual distances are given by
\begin{align*}
l_{23}(0) &= \dfrac{l_{12}(0)}{| \Gamma_1 + \Gamma_2 |} \sqrt{ \Gamma_1^2 + \mathcal{R} + 2 \Gamma_1 \sqrt{\mathcal{R}} \cos{\theta} }, \\
l_{31}(0) &= \dfrac{l_{12}(0)}{| \Gamma_1 + \Gamma_2 |} \sqrt{\Gamma_2^2 + \mathcal{R} - 2 \Gamma_2 \sqrt{\mathcal{R}} \cos{\theta} }, \\
l_{12}(0) &= \frac{\sqrt{\mathcal{R}}}{|\Gamma_1 + \Gamma_2|},
\end{align*}
and three point vortices form equilateral triangles for $\theta = \theta_\pm$ satisfying $\cos{\theta_\pm} = - (\Gamma_1 - \Gamma_2)/(2 \sqrt{\mathcal{R}})$ and $0 < \theta_+ < \pi < \theta_- < 2\pi$.
Since the initial configuration \eqref{init:three} are a one-parameter family with the parameter $\theta \in [0,2 \pi)$, $\mathscr{H}$ and $(A, B)$ in \eqref{def-AB} are considered as functions of $\theta$:
\begin{align}
\mathscr{H}(\theta) &= \dfrac{\Gamma_1 \Gamma_2}{4\pi (\Gamma_1 + \Gamma_2)} \log{\left( \dfrac{\Gamma_1^2 + \mathcal{R} + 2 \Gamma_1 \sqrt{\mathcal{R}} \cos{\theta}}{(\Gamma_1 + \Gamma_2)^2} \right)^{\Gamma_2} \left( \dfrac{\Gamma_2^2 + \mathcal{R} - 2 \Gamma_2 \sqrt{\mathcal{R}} \cos{\theta}}{(\Gamma_1 + \Gamma_2)^2} \right)^{\Gamma_1}}, \nonumber \\
A(\theta) &= \sigma_{123} \dfrac{\Gamma_1 \Gamma_2 |\Gamma_1 + \Gamma_2|^3}{2 \pi \sqrt{\mathcal{R}}}  \dfrac{\sqrt{1 - \cos^2{\theta}} (\Gamma_1 - \Gamma_2 + 2 \sqrt{R} \cos{\theta}) }{(\Gamma_1^2 + \mathcal{R} + 2 \Gamma_1 \sqrt{\mathcal{R}} \cos{\theta}) (\Gamma_2^2 + \mathcal{R} - 2 \Gamma_2 \sqrt{\mathcal{R}} \cos{\theta})}, \label{def:A-three} \\
B(\theta) &= \dfrac{(\Gamma_1 + \Gamma_2)^3}{2 \pi \mathcal{R}} \left[ 1 + \Gamma_1 \Gamma_2 \dfrac{ (\Gamma_1 - \Gamma_2) \sqrt{\mathcal{R}}\cos{\theta} + 2 \mathcal{R} \cos^2{\theta}  - (\Gamma_1 + \Gamma_2)^2 }{(\Gamma_1^2 + \mathcal{R} + 2 \Gamma_1 \sqrt{\mathcal{R}} \cos{\theta}) (\Gamma_2^2 + \mathcal{R} - 2 \Gamma_2 \sqrt{\mathcal{R}} \cos{\theta})} \right]. \nonumber
\end{align}
All possible equilibria, which are equivalent to $A = 0$, are collinear states for $\theta = 0, \pi$ and equilateral triangles for $\theta = \theta_\pm$. It is easily confirmed that the self-similar solution is collapsing for $\theta \in (0, \theta_+) \cup (\pi, \theta_-)$ and expanding for $\theta \in (\theta_+, \pi) \cup (\theta_-, 2 \pi)$. In the case of $\Gamma_1 = \Gamma_2$, which is the case we use for numerical computations later, we have $\theta_+ = \pi/2$ and $\theta_- = 3 \pi/2$.

In the PV system with $N \geq 4$, explicit formulae for configurations leading to self-similar collapse have not been established in general. On the other hand, an example of exact self-similar collapsing solutions for the four and five vortex problems has been obtained in \cite{Novikov}. In this example, four point vortices are located at vertices of a parallelogram whose diagonals intersect at the origin and the fifth point vortex is located at the origin, that is,
\begin{equation}
k_1 = \dfrac{1}{2} d_1 e^{i\theta}, \qquad k_2 = - \dfrac{1}{2} d_1 e^{i\theta}, \qquad k_3 = - \dfrac{1}{2} d_2, \qquad k_4 = \dfrac{1}{2} d_2, \qquad k_5 = 0, \label{init:Novikov}
\end{equation}
where given constants $d_1, d_2$ are lengths of the diagonals and $\theta \in [0,2\pi)$ is the angle between the diagonals. The strengths of point vortices are $\Gamma_1 = \Gamma_2 = \alpha$, $\Gamma_3 = \Gamma_4 = \beta$ and $\Gamma_5 = \gamma$ where $\alpha, \beta, \gamma \in \mathbb{R} \setminus \{ 0 \}$ are given constants. The configuration \eqref{init:Novikov} satisfies $P = Q = 0$ and, due to the self-similarity with rotation, it should do
\begin{align*}
& I = \dfrac{1}{2}\left( \alpha d_1^2 + \beta d_2^2 \right) = 0, \\
& M = \dfrac{1}{2}(2 \alpha + 2 \beta + \gamma)\left( \alpha d_1^2 + \beta d_2^2 \right) = 0, \\
& \Gamma_H = \alpha^2 + 4 \alpha \beta + \beta^2 + 2 \gamma (\alpha + \beta)= 0
\end{align*}
and these conditions yield the relation $d_1^2 / d_2^2 = - \beta/\alpha$. Ignoring the fifth point vortex $k_5$, we have the initial configuration leading to self-similar collapse for the four PV system. Similarly to the five PV problem, we have $P = Q = 0$ and the following conditions should be satisfied.
\begin{align*}
& I = \dfrac{1}{2}\left( \alpha d_1^2 + \beta d_2^2 \right) = 0,\\
& M = (\alpha + \beta)\left( \alpha d_1^2 + \beta d_2^2 \right) = 0, \\
& \Gamma_H = \alpha^2 + 4 \alpha \beta + \beta^2 = 0.
\end{align*}
Then, we find $d_1^2 / d_2^2 = - \beta/\alpha = 2 \pm \sqrt{3} > 0$ and thus $\alpha$ and $\beta$ have opposite signs. Considering \eqref{init:Novikov} as an one-parameter family, we obtain formulae for $\mathscr{H}$ and $(A,B)$ that are functions of $\theta$:
\begin{align}
\mathscr{H}(\theta) &= \dfrac{-1}{2 \pi} \log{\left[ c_H d_1^{\alpha (\alpha + 2 \gamma)} d_2^{\beta (\beta + 2 \gamma)} \left(d_1^4 + d_2^4 - 2 d_1^2 d_2^2 \cos{2 \theta} \right)^{\alpha \beta} \right]}, \nonumber \\
A(\theta) &= \dfrac{4 \alpha d_1^2 \sin{2 \theta}}{\pi \left(d_1^4 + d_2^4 -2 d_1^2 d_2^2 \cos{2 \theta}  \right)}, \label{def:A-Nov} \\
B(\theta) &= \dfrac{d_1^2 + d_2^2}{2 \pi d_1^2 d_2^2} \left[ \alpha + \beta + 2 \gamma + \dfrac{4 \beta d_2^2(d_1^2 - d_2^2)}{d_1^4 + d_2^4 -2 d_1^2 d_2^2 \cos{2 \theta}} \right], \nonumber
\end{align}
where $c_H := 2^{ -4 \alpha \beta - 2 \gamma (\alpha + \beta)}$. These formulae are valid for the four vortex problem by setting $\gamma \equiv 0$, see \cite{Novikov}. In both the four and five PV system, relative equilibria are diamond configurations for $\theta = \pi/2, 3\pi/2$ and collinear states for $\theta = 0, \pi$.
In this paper, we consider the case of $\alpha < 0$, for which the self-similar motions are collapsing for $\theta \in (0, \pi/2) \cup (\pi, 3\pi/2)$ and expanding for $\theta \in (\pi/2, \pi) \cup (3\pi/2, 2\pi)$.

\section{The filtered-point vortex system}
\label{sec:FPVS}

\subsection{Dynamics of point vortices on the 2D filtered-Euler flow}
\label{subsec:introduction}

We introduce the filtered-point-vortex (FPV) system, which describes the dynamics of point-vortex solutions of the 2D filtered-Euler equations. The filtered-Euler equations are a regularized model of the Euler equations on the basis of a spatial filtering and given by
\begin{equation}
\partial_t \boldsymbol{v}^\eps + (\boldsymbol{u}^\varepsilon\cdot\nabla)\boldsymbol{v}^\eps - (\nabla\boldsymbol{v}^\eps)^T \cdot \boldsymbol{u}^\varepsilon - \nabla p^\varepsilon = 0, \quad \bs{u}^\eps = h^\eps \ast \bs{v}^\eps, \quad \nabla \cdot \boldsymbol{v}^\eps = 0, \label{eq:FE}
\end{equation}
where $\bs{v}^\eps$ and $p^\eps$ are unknown functions and $h^\eps$ has the form
\begin{equation*}
h^\varepsilon(\boldsymbol{x}) := \frac{1}{\varepsilon^2} h \left( \frac{|\boldsymbol{x}|}{\varepsilon} \right)
\end{equation*}
with a given filter function $h \in L^1(0,\infty)$ \cite{Foias, Holm}. We mention detailed properties that a filter function $h$ should satisfy in Remark~\ref{rem:h}. We consider the vorticity $q^\eps := \operatorname{curl} \bs{v}^\eps$ and the vorticity form of \eqref{eq:FE}:
\begin{equation}
\partial_t q^\eps + (\boldsymbol{u}^\varepsilon \cdot \nabla)q^\eps = 0,  \qquad
\boldsymbol{u}^\varepsilon = \boldsymbol{K}^\varepsilon \ast q^\eps, \label{eq:VFE}
\end{equation}
where $\boldsymbol{K}^\varepsilon := \boldsymbol{K} \ast h^\varepsilon$ is a filtered kernel and we call the relation $\boldsymbol{u}^\varepsilon = \boldsymbol{K}^\varepsilon \ast q^\eps$ the filtered-Biot-Savart law. In contrast to the 2D Euler equations, the 2D filtered-Euler equations have a unique global weak solution for initial vorticity $q_0 \in \mathcal{M}(\R^2)$ \cite{G1}, which guarantees the global solvability for the point-vortex initial data \eqref{PVI}. Thus, we have the filtered-Lagrangian flow map $\bs{\eta}^\eps$ convected by the filtered velocity $\bs{u}^\eps$:
\begin{equation}
\partial_t \boldsymbol{\eta}^\varepsilon(\boldsymbol{x}, t) = \boldsymbol{u}^\varepsilon \left( \boldsymbol{\eta}^\varepsilon(\boldsymbol{x}, t), t \right), \qquad  \boldsymbol{\eta}^\varepsilon(\boldsymbol{x}, 0) = \boldsymbol{x} \label{eq:FMFE}
\end{equation}
and the solution of \eqref{eq:VFE} is expressed by $q^\eps(\bs{x},t) = q_0(\bs{\eta}^\eps(\bs{x}, -t))$.
The convergence of the 2D filtered-Euler equations to the 2D Euler equations has been shown for the initial vorticity in $L^1(\R^2) \cap L^p(\R^2)$ with $1 < p \leq \infty$ and $\mathcal{M}(\mathbb{R}^2) \cap H_{\mathrm{loc}}^{-1}(\mathbb{R}^2)$, see \cite{G1,G2}.

\begin{remark}
The singular kernel $\bs{K}$ appeared in the 2D Euler equations is expressed by $\boldsymbol{K} = \nabla^\perp G$, where $\nabla^\perp = (- \partial_{x_2}, \partial_{x_1})$ and $G$ is a fundamental solution to the 2D Laplacian $\Delta G = \delta $. On the other hand, the filtered-kernel $\boldsymbol{K}^\varepsilon$ in the 2D filtered-Euler equations is represented by $\boldsymbol{K}^\varepsilon = \nabla^\perp G^\varepsilon$, where $G^\eps$ is a solution to the 2D Poisson equation $\Delta G^\varepsilon = h^\varepsilon$.
Since $h^\eps$ is radially symmetric, it is easily confirmed that $\bs{K}^\eps(\bs{x}) = \nabla^\perp \left(G_r (|\bs{x}|/ \eps) \right)$ and $G_r$ satisfies
\begin{equation*}
\frac{1}{r} \frac{\mbox{d}}{\mbox{d}r} \left( r  \frac{\mbox{d}}{\mbox{d}r} G_r(r) \right) = h(r).
\end{equation*}
Then, we have
\begin{equation}
\boldsymbol{K}^\varepsilon(\boldsymbol{x}) = \boldsymbol{K}(\boldsymbol{x})\, P\left( \frac{|\boldsymbol{x}|}{\varepsilon} \right), \qquad P(r) := - 2\pi r \frac{\mbox{d} G_r}{\mbox{d}r}(r). \label{def:P}
\end{equation}
The function $P \in C[0,\infty)$ is monotonically increasing and satisfies $P(0)=0$ and $P(r) \to  1$ as $r \to \infty$. Note that $\bs{K}^\eps$ does not have singularity at the origin and belongs to $C_0(\R^2)$, the space of continuous functions vanishing at infinity.
\end{remark}

We consider point-vortex solutions of \eqref{eq:VFE}. Owing to the uniqueness for the initial vorticity \eqref{PVI}, we have a unique filtered-Lagrangian flow map $\bs{\eta}^\eps$ and the solution of \eqref{eq:VFE} is given by
\begin{equation*}
q^\eps(\boldsymbol{x},t) = \sum_{m=1}^N \Gamma_m \ \delta(\boldsymbol{x} - \bs{\eta}^\eps(\bs{k}_m,t)).
\end{equation*}
Setting $\bs{x}^\eps_m(t) := \bs{\eta}^\eps(\bs{k}_m,t)$, we find from \eqref{eq:FMFE} and the filtered-Biot-Savart law that
\begin{equation*}
\frac{\mbox{d}}{\mbox{d}t} \bs{x}^\eps_m(t) = \sum_{n=1}^N \Gamma_n \int \bs{K}^\eps ( \bs{x}^\eps_m(t) - \bs{y})\ \delta(\bs{y} - \bs{x}^\eps_n(t)) d\bs{y} = \sum_{n\neq m}^N \Gamma_n \bs{K}^\eps (\bs{x}^\eps_m(t) - \bs{x}^\eps_n(t)).
\end{equation*}
Let $z_m^\eps(t) := x_m^\eps(t) + i y_m^\eps(t)$ and recall \eqref{def:P}. Then, we obtain the FPV system in the complex form:
\begin{equation}
\dfrac{\mbox{d} z_m^\eps}{\mbox{d}t} = \frac{-1}{2 \pi i} \sum_{n\neq m}^N \dfrac{\Gamma_n}{\overline{z_m^\eps} - \overline{z_n^\eps}} P\left( \frac{l_{mn}^\eps}{\eps} \right), \qquad z_m^\eps(0) = k_m \label{eq:FPV}
\end{equation}
for $m = 1,\cdots, N$, where $l_{mn}^\eps(t) := |z_m^\eps(t) - z_n^\eps(t)|$, see \cite{GS3}. We call point vortices governed by the FPV system {\it filtered-point vortices}. The FPV system is a Hamiltonian dynamical system satisfying
\begin{equation*}
\Gamma_m \frac{\mbox{d} x_m^\varepsilon}{\mbox{d}t}  = \frac{\partial \mathscr{H}^\varepsilon}{\partial y_m^\varepsilon}, \qquad \Gamma_m \frac{\mbox{d} y_m^\varepsilon}{\mbox{d}t}  = - \frac{\partial \mathscr{H}^\varepsilon}{\partial x_m^\varepsilon}
\end{equation*}
for $m = 1,\dots,N$ with the Hamiltonian
\begin{equation*}
\mathscr{H}^\varepsilon := - \frac{1}{2 \pi} \sum_{m=1}^N \sum_{n=m+1}^N \Gamma_m \Gamma_n \left[ \log{l_{mn}^\varepsilon} + H_G\left( \frac{l_{mn}^\varepsilon}{\varepsilon} \right) \right],
\end{equation*}
where $H_G(r) := - \log{r} - 2 \pi G_r(r)$. Similarly to the PV system, the FPV system admits the conserved quantities,
\begin{equation*}
P^\varepsilon + i Q^\varepsilon := \sum_{m = 1}^N \Gamma_m x_m^\varepsilon + i \sum_{m = 1}^N \Gamma_m y_m^\varepsilon, \quad I^\varepsilon := \sum_{m = 1}^N \Gamma_m |z_m^\varepsilon|^2, \quad M^\varepsilon := \sum_{m = 1}^N \sum_{n = m+1}^N \Gamma_m \Gamma_n (l_{mn}^\varepsilon)^2
\end{equation*}
and the Hamiltonian $\mathscr{H}^\varepsilon$. The integrability depending on $N$ is the same as the PV system. The evolution of mutual distances $l^\eps_{mn}$ is governed by
\begin{equation*}
\frac{\mbox{d}}{\mbox{d} t} \left( l^\eps_{mn} \right)^2 = \frac{2}{\pi} \sum_{l \neq n \neq m}^N \Gamma_l \sigma^\eps_{mnl} A^\eps_{mnl} \left[ \frac{1}{\left(l^\eps_{nl}\right)^2} P\left( \frac{l^\eps_{nl}}{\eps} \right) - \frac{1}{\left(l^\eps_{lm} \right)^2} P\left( \frac{l^\eps_{lm}}{\eps} \right) \right],  \label{eq:lmn-FPV}
\end{equation*}
where $\sigma^\eps_{mnl}$ denotes the sign of the area and $A^\eps_{mnl}$ does the area of the triangle. It is important to remark that, owing to the global solvability and uniqueness, filtered-point vortices never collapse for any $\varepsilon > 0$. In contrast to the 2D Euler equations, the derivation of \eqref{eq:FPV} is mathematically rigorous and thus the FPV system is equivalent to the vorticity form of the 2D filtered Euler equations with initial data \eqref{PVI}: a weak solution to the 2D filtered-Euler equations yields a solution to the FPV system and vice versa.

\begin{remark} \label{rem:h}
The filter function $h$ characterizes the regularity of the filtered-model. In this paper, we suppose that $h$ is a given function satisfying $h \in C_0(0, \infty) \cap L^1(0,\infty)$ and
\begin{equation*}
2 \pi \int_0^\infty r h (r) dr = 1.
\end{equation*}
Note that $h$ may have a singularity at the origin but should decay at infinity, see \cite{G1,GS3} for the detailed condition that guarantees the global solvability for the point-vortex initial vorticity. For a suitable filter function $h$, the 2D filtered-Euler equations have a unique global weak solution for initial vorticity in $\mathcal{M}(\R^2)$. Considering a specific filter function $h$, we obtain an explicit form of the filtered-Euler equations. For instance, the Euler-$\alpha$ model, the vortex blob regularization and the exponential model are often used as a regularized inviscid model.
\end{remark}

\subsection{Variations of enstrophy and energy}
\label{subsec:variations}

We introduce variations of the enstrophy and the energy for solutions to the FPV system. The derivations of them are based on the Fourier transform and we here start with the final forms of the total enstrophy and the approximated total energy, see \cite{GS3} for the detailed derivations. We define the total enstrophy by the $L^2$-norm of the filtered vorticity $\omega^\eps:= h^\eps \ast q^\eps$. Then, the total enstrophy of the FPV system is given by
\begin{align*}
\frac{1}{2}\int_{\mathbb{R}^2} \left\vert \omega^\varepsilon({\boldsymbol x}, t) \right\vert^2 d{\boldsymbol x} &= \frac{1}{4 \pi \varepsilon^2} \sum_{m=1}^N \Gamma_m^2 \int_0^\infty s \left\vert 2\pi\widehat{h}(s) \right\vert^2 ds \\
&\quad + \frac{1}{2\pi \varepsilon^2} \sum_{m=1}^N \sum_{n=m+1}^N \Gamma_m \Gamma_n \int_0^\infty s \left\vert 2\pi\widehat{h}(s) \right\vert^2  J_0 \left( s \frac{l_{mn}^\varepsilon}{\varepsilon} \right) ds,
\end{align*}
where $\widehat{h}$ is the Fourier transform of $h$ and $J_0$ is a Bessel function of the first kind. Since the first term in the right-hand side is constant in time, a variational part of the total enstrophy is provided by the second term,
\begin{equation*}
\mathscr{Z}^\varepsilon(t) := \frac{1}{2\pi \varepsilon^2} \sum_{m=1}^N \sum_{n=m+1}^N \Gamma_m \Gamma_n \int_0^\infty s \left\vert 2\pi\widehat{h}(s) \right\vert^2  J_0 \left( s \frac{l_{mn}^\varepsilon(t)}{\varepsilon} \right) ds.
\end{equation*}
In this paper, we call the variation $\mathscr{Z}^\varepsilon$ the {\it enstrophy} of the FPV system.
The total energy for the filtered-model is defined by the $L^2$-norm of the filtered velocity $\bs{u}^\eps$. However, the total energy on the whole space $\R^2$ is not finite in general, since the filtered-Biot-Savart law implies $\bs{u}^\eps(\bs{x}) \sim |\bs{x}|^{-1}$ as $|\bs{x}| \to \infty$. Thus, we consider the total energy cut off at a scale larger than $L \gg 1$. Then, the following approximation holds.
\begin{align*}
\frac{1}{2} & \int_{\mathbb{R}^2} \left\vert {\boldsymbol u}^\varepsilon({\boldsymbol x}, t) \right\vert^2 d{\boldsymbol x} \sim \frac{1}{4 \pi} \sum_{m=1}^N \Gamma_m^2\int_{\varepsilon L^{-1}}^\infty \frac{1}{s} \left\vert 2\pi\widehat{h}(s) \right\vert^2 ds \\
& \quad + \frac{1}{2\pi} \sum_{m=1}^N \sum_{n=m+1}^N \Gamma_m \Gamma_n \left( \log{\frac{L e^\zeta}{2}} + \mathcal{O}\left( L^{-2} \log{L^{-1}} \right) \right) \\
& \quad - \frac{1}{2\pi} \sum_{m=1}^N \sum_{n=m+1}^N \Gamma_m \Gamma_n \left[ \log{l_{mn}^\varepsilon(t)} +  \int_{\varepsilon L^{-1}}^\infty \frac{1}{s} \left( 1 - \left\vert 2\pi\widehat{h}(s) \right\vert^2 \right) J_0 \left( s \frac{l_{mn}^\varepsilon(t)}{\varepsilon} \right) ds \right],
\end{align*}
where $\zeta$ is the Euler's constant. Taking the $L \rightarrow \infty$ limit in the non-constant part, we obtain a variational part of the approximated total energy:
\begin{equation*}
\mathscr{E}^\eps(t) := - \frac{1}{2\pi} \sum_{m=1}^N \sum_{n=m+1}^N \Gamma_m \Gamma_n \left[ \log{l_{mn}^\varepsilon(t)} +  \int_0^\infty \frac{1}{s} \left( 1 - \left\vert 2\pi\widehat{h}(s) \right\vert^2 \right) J_0 \left( s \frac{l_{mn}^\varepsilon(t)}{\varepsilon} \right) ds \right],
\end{equation*}
in which the integrand of the second term does not have any singularity owing to $2\pi \widehat{h}(0) = 1$. Then, we define the {\it energy dissipation rate} of the FPV system by the time derivative of $\mathscr{E}^\eps(t)$:
\begin{equation*}
\mathscr{D}_E^\eps (t) := \frac{\mbox{d}}{\mbox{d} t} \mathscr{E}^\eps(t).
\end{equation*}

\subsection{Preceding results about enstrophy dissipation}
\label{subsec:Preceding}

As the first attempt of constructing a point-vortex solution dissipating the enstrophy, it has been numerically shown in \cite{Sakajo} that self-similar collapse of three point vortices could dissipate the enstrophy by using the Euler-$\alpha$ model. Let us review the preceding results for the Euler-$\alpha$ model more precisely since we use this model later for numerical computations. In the Euler-$\alpha$ model, the filter function $h$ is given by
\begin{equation}
h(r) = \frac{1}{2 \pi} K_0(r), \label{def:h-alpha}
\end{equation}
where $K_0$ is the $0$-th order modified Bessel function of the second kind, for which the corresponding equations \eqref{eq:FE} are called the 2D Euler-$\alpha$ equations.
Then, we find that the smoothing function $P$ in the FPV system \eqref{eq:FPV} is expressed by
\begin{equation}
P(r) = 1 - r K_1(r), \label{def:P-alpha}
\end{equation}
where $K_1$ denotes the $1$-st order modified Bessel function of the second kind, see \cite{GS1, GS2, Sakajo}. The FPV system and filtered-point vortices with \eqref{def:P-alpha} are called the {\it $\alpha$-point-vortex ($\alpha$PV) system} and {\it $\alpha$-point vortices}. Note that the filter parameter in the Euler-$\alpha$ model is often denoted by $\alpha$, but we consistently use $\varepsilon$ to avoid confusion in this paper.
The quantities $\mathscr{H}^\eps$, $\mathscr{Z}^\eps$ and $\mathscr{E}^\eps$ for the $\alpha$PV system are explicitly described by
\begin{align*}
\mathscr{H}^\varepsilon &= - \frac{1}{2 \pi} \sum_{m=1}^N \sum_{n=m+1}^N \Gamma_m \Gamma_n \left[ \log{l_{mn}^\varepsilon} + K_0\left( \frac{l_{mn}^\varepsilon}{\varepsilon} \right) \right], \\
\mathscr{Z}^\varepsilon &= \frac{1}{4\pi \varepsilon^2} \sum_{m=1}^N \sum_{n=m+1}^N \Gamma_m \Gamma_n \frac{l_{mn}^\varepsilon}{\varepsilon} K_1 \left( \frac{l_{mn}^\varepsilon}{\varepsilon} \right), \\
\mathscr{E}^\eps &= - \frac{1}{2\pi} \sum_{m=1}^N \sum_{n=m+1}^N \Gamma_m \Gamma_n \left[ \log{l_{mn}^\varepsilon} + K_0 \left( \frac{l_{mn}^\varepsilon}{\varepsilon} \right) + \frac{l_{mn}^\varepsilon}{2 \varepsilon} K_1 \left( \frac{l_{mn}^\varepsilon}{\varepsilon} \right) \right],
\end{align*}
and they satisfy $\mathscr{E}^\eps + \eps^2 \mathscr{Z}^\varepsilon = \mathscr{H}^\varepsilon$. The preceding study \cite{Sakajo} has considered the three $\alpha$PV system and shown that, under the condition $\Gamma_H = M ^\eps = 0$, three $\alpha$-point vortices converge to a self-similar collapsing solution of the PV system in the $\eps \to 0$ limit and dissipate the enstrophy at the event of the triple collapse while the energy is conserved.
These results have been shown with a help of numerical computations in \cite{Sakajo} and subsequently proven with a mathematical rigor in \cite{GS2}. Then, the study \cite{GS3} has proven that the same results in \cite{GS2,Sakajo} hold for the FPV system \eqref{eq:FPV} with the general filter function $h$. In these preceding studies, the enstrophy dissipation by collapse of point vortices and the energy conservation mean that there exist constants $t_c \in \R$ and $m_z < 0$ such that we have
\begin{equation}
\lim_{\varepsilon \rightarrow 0} l^\varepsilon_{mn}(t_c) = 0 \label{thm:lmn}
\end{equation}
and
\begin{equation}
\lim_{\varepsilon \rightarrow 0} \mathscr{Z}^\varepsilon = m_z \delta(\cdot - t_c) , \qquad \lim_{\varepsilon \rightarrow 0} \mathscr{D}_E^\varepsilon = 0 \label{thm:Z-DE}
\end{equation}
in the sense of distributions. Here, $t_c$ is the time when the self-similar collapse occurs and $m_z$ is the mass of the enstrophy dissipation. For the $N$ vortex problem with $N \geq 4$, we have one numerical example of a quadruple collapse causing the enstrophy dissipation in the four $\alpha$PV system \cite{GS1} with a specific initial data that is also found numerically. In the present study, we consider the four and five vortex problem in the FPV system with initial data \eqref{init:Novikov} and give numerical solutions leading to the enstrophy dissipation by the vortex collapse in the $\eps \to 0$ limit.

\section{Main results}
\label{sec:main}

\subsection{Numerical method}
\label{subsec:methods}

In order to conduct numerical computations for dynamics of solutions to the FPV system, we have to give an explicit filter function $h$.
In this paper, we employ the $\alpha$PV system, that is, the FPV system \eqref{eq:FPV} with \eqref{def:h-alpha}, and consider the three, four and five vortex problems: the three $\alpha$PV system with \eqref{init:three}, the four $\alpha$PV system with \eqref{init:Novikov} without $k_5$ and the five $\alpha$PV system with \eqref{init:Novikov}.
Similarly to the PV system, solutions to the FPV system for $\theta \in (\pi, 2\pi)$ are symmetric to those for $\theta \in (0, \pi)$ with the real axis. In addition, solutions for $\theta \in (\pi/2, \pi)$ are expanding and thus there is no collapse of filtered-point vortices in the $\eps \to 0$ limit for any positive time.
For the cases of $\theta = 0, \pi/2, \pi$ and $3\pi/2$ on the axes, as we mentioned in Section~\ref{subsec:exact-solutions}, the corresponding solutions to the PV system are relative equilibria. However, the solutions to the FPV system for those initial configurations are not relative equilibria and they are expanding, expect for the three filtered-point vortices with $\theta = \pi/2$. Thus, it is enough to pay attention to $\theta \in (0, \pi/2)$ for considering the enstrophy variations by vortex collapse.

For later use, we introduce several notations. Since solutions to the FPV system are parametrized by $\theta$ owing to the initial data \eqref{init:three} and \eqref{init:Novikov}, we describe
\begin{equation*}
l_{mn}^\eps(t;\theta) = l_{mn}^\eps(t)
\end{equation*}
when we emphasize that the solutions depend on $\theta$. Then, for $\theta \in (0,\pi/2)$, we define the critical time $t_c^\eps = t_c^\eps(\theta)$ as the time when the total length of $l_{mn}^\eps$ in the $L^2$-sense,
\begin{equation*}
L^\eps(t; \theta) := \sum _{m =1}^N \sum_{n = m + 1}^N \left( l^\eps_{mn}(t;\theta) \right)^2,
\end{equation*}
attains its minimum, that is,
\begin{equation*}
t_c^\eps = t_c^\eps(\theta) := \underset{t \geq 0}{\operatorname{argmin}}\ L^\eps(t; \theta).
\end{equation*}
The critical time $t_c^\eps$ corresponds to the collapse time of the PV system. Indeed, for initial data \eqref{init:three} and \eqref{init:Novikov}, the collapse time \eqref{def:tc} of the PV system is a function of $\theta$, which we describe by
\begin{equation}
t_c = t_c(\theta) := \frac{1}{2 A(\theta)}, \label{def:tc-theta}
\end{equation}
where $A(\theta)$ is given by \eqref{def:A-three} or \eqref{def:A-Nov}. Then, as we see later, numerical computations show that $t_c^\eps(\theta)$ converges to $t_c(\theta)$ in the $\eps \to 0$ limit for any $\theta \in (0, \pi/2)$.
The enstrophy $\mathscr{Z}^\eps$ and the energy dissipation rate $\mathscr{D}_E^\eps$ are also functions of $\theta$ and denoted by $\mathscr{Z}^\eps(t; \theta)$ and $\mathscr{D}^\eps_E(t; \theta)$. We represent the values of $L^\eps(t; \theta)$ and $\mathscr{Z}^\eps(t; \theta)$ at the critical time by
\begin{equation*}
L^\eps_c(\theta) := L^\eps(t_c^\eps(\theta), \theta), \qquad \mathscr{Z}^\eps_c(\theta) := \mathscr{Z}^\eps(t_c^\eps(\theta), \theta)
\end{equation*}
for simplicity. We remark that the convergence of $L^\eps_c(\theta)$ to zero and the divergence of $\mathscr{Z}^\eps_c(\theta)$ to negative infinity in the $\eps \to 0$ limit indicate the collapse of point vortices and the enstrophy dissipation, respectively.

\begin{figure}[t]
\begin{center}
\includegraphics[scale=0.7]{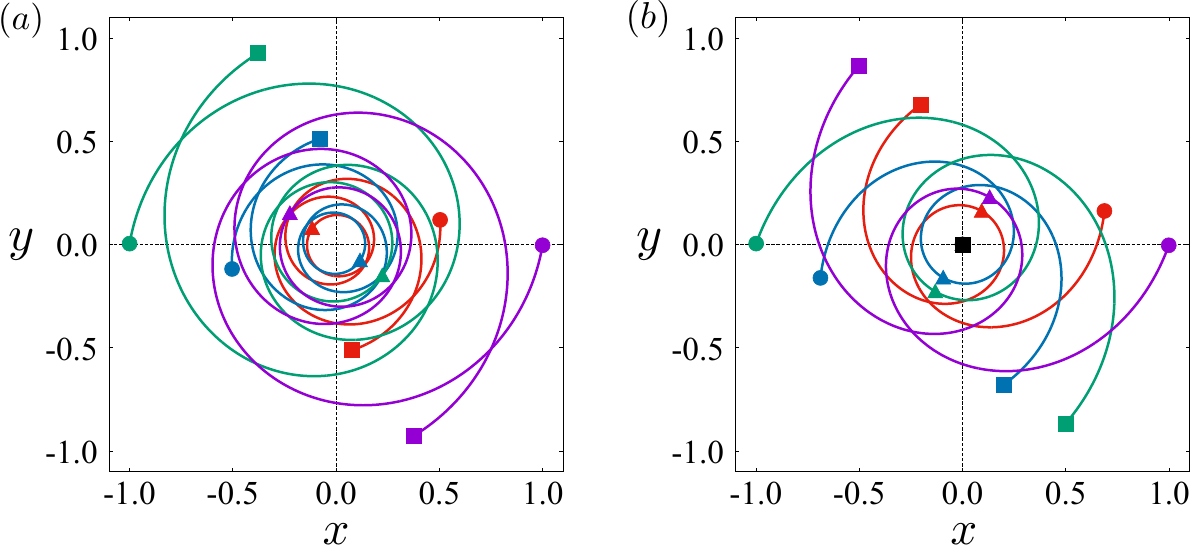}
\caption{The orbits of the (a) four and (b) five $\alpha$-point vortices for the initial data \eqref{init:Novikov} with $\theta = \theta_{30}$ and the filter parameter $\eps_3 = 0.05$. The circular, triangle and square points describe the configurations at $t=0$, $t=t_c^\eps$ and $t=2t_c^\eps$, respectively.}
\label{fig:fpv_orbits_N4N5}
\end{center}
\end{figure}

For numerical computations, we divide the interval $(0, \pi/2)$ into $200$ segments and compute solutions to the $\alpha$PV system with the initial configuration for
\begin{equation*}
\theta_i := \frac{\pi}{2} \times \frac{i}{200}, \qquad i = 1, \cdots, 199.
\end{equation*}
As for the filter parameter $\eps$, we compute the five cases of
\begin{equation}
\eps_1 := 0.01, \quad \eps_2 := 0.025, \quad \eps_3 := 0.05, \quad \eps_4 := 0.075, \quad \eps_5 := 0.1. \label{def:eps-n}
\end{equation}
As the numerical scheme for solving the $\alpha$PV systems, we use the 5-stage implicit Runge-Kutta method based on the n-point Gauss-Legendre quadrature formula \cite{Butcher} with the time step size $\Delta t = 0.0001$. To ensure the accuracy of numerical solutions, we use variables with 32 decimal digit precision. For the four $\alpha$PV system with several initial configurations near $\theta = 0$, we have used $\Delta t = 0.00001$ and 50 decimal digit precision to accurately compute long time behaviors of the solutions. For the five $\alpha$PV system, mathematical analysis shows that the fifth point vortex is fixed to the origin throughout time evolution and thus we have fixed it in the numerical scheme. Figure~\ref{fig:fpv_orbits_N4N5} shows examples of the four and five $\alpha$-point vortices for the initial data \eqref{init:Novikov} with $\theta = \theta_{30}$.

\subsection{Three vortex problem}
\label{subsec:results-three}

\begin{figure}[t]
\begin{center}
\includegraphics[scale=0.70]{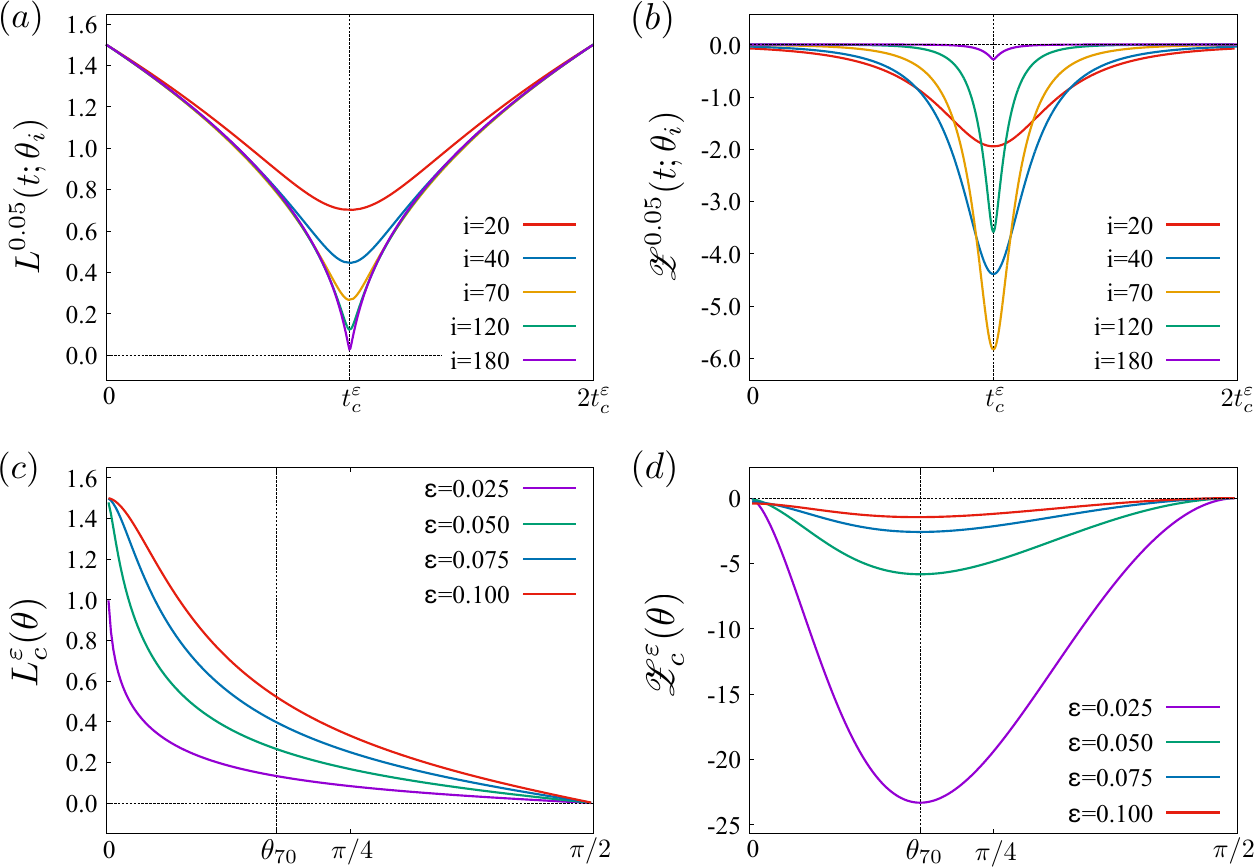}

\caption{The graphs of (a) $\{ L^{\eps_3}(t; \theta_i) \}_{i \in \mathcal{I}}$, (b) $\{ \mathscr{Z}^{\eps_3}(t;\theta_i) \}_{i \in \mathcal{I}}$ with $\mathcal{I}= \{20, 40, 70, 120, 180 \}$,
(c) $\{ L^{\eps_n}_c(\theta)\}_{n=2}^5$ and (d) $\{ \mathscr{Z}^{\eps_n}_c(\theta) \}_{n=2}^5$ for the three $\alpha$PV system. In (a) and (b), the time axes are rescaled so that $\{ t_c^\eps(\theta_i) \}_{i \in \mathcal{I}}$ are placed at the same midpoint. The curves in (c) and (d) are interpolating data for $i = 1, \cdots, 199$ with lines.}

\label{fig:N3_alpha_LZ}
\end{center}
\end{figure}

In the three vortex problem, it has already been proven in \cite{GS2, GS3} that the solution to the FPV system with initial data \eqref{init:three} for $\theta \in (0, \pi/2)$ converges to a self-similar collapsing solution of the PV system and dissipates the enstrophy at the collapse time.
In order to see the detailed processes of the vortex collapse and the enstrophy dissipation, we investigate the three $\alpha$PV system with initial data \eqref{init:three} by numerical computations. For simplicity, we consider the case of $\Gamma_1 = \Gamma_2$: we may set $\Gamma_1 = \Gamma_2 = -1$ and $\Gamma_3 = 1/2$ without loss of generality. Then, \eqref{init:three} is expressed by
\begin{equation}
k_1 = \dfrac{1}{4}\left( 1 + \sqrt{3} e^{-i\theta} \right), \qquad  k_2 = \dfrac{1}{4}\left( 1 - \sqrt{3} e^{-i\theta} \right), \qquad k_3 = 1 \label{init:three-smpl}
\end{equation}
for $\theta \in (0, \pi/2)$.
Figure~\ref{fig:N3_alpha_LZ}~(a) and (b) show the graphs of $L^{\eps_3}(t;\theta_i)$ and $\mathscr{Z}^{\eps_3}(t; \theta_i)$ for $i = 20, 40, 70, 120, 180$.
These graphs indicate that, for any fixed $\theta \in (0, \pi/2)$, the functions $L^\eps(t;\theta)$ and $\mathscr{Z}^\eps(t;\theta)$ of the variable $t$ are monotonically decreasing for $t < t_c^\eps$ and increasing for $t > t_c^\eps$: $\mathscr{Z}^\eps(t;\theta)$ attains its minimum at $t_c^\eps(\theta)$ that is defined by the time when $L^\eps(t;\theta)$ reaches its minimum.
That is to say, three $\alpha$-point vortices approach each other with decreasing the enstrophy as time evolves, and then the enstrophy decreases the most when $\alpha$-point vortices are at their closest. After the critical time, the enstrophy increases as the $\alpha$-point vortices move away from each other.
Note that $\mathscr{Z}^\eps(t;\theta)$ is a negative function for any fixed $\theta \in (0, \pi/2)$, which is specific to the three vortex problem as we see in the four and five vortex problems later.

We focus on the total length and the enstrophy at the critical time. As we see in Figure~\ref{fig:N3_alpha_LZ}~(c), $L^\eps_c(\theta)$ is monotonically decreasing and, for any fixed $\theta \in (0,\pi/2)$, it seems to approach zero as $\eps$ gets smaller.
On the other hand, Figure~\ref{fig:N3_alpha_LZ}~(d) shows that $\mathscr{Z}^\eps_c(\theta)$ has the minimum in $(0,\pi/2)$ and numerical computations suggest
\begin{equation*}
\theta_{69} < \operatorname{argmin}_{\theta \in (0,\pi/2)}\, \mathscr{Z}^\eps_c(\theta) < \theta_{71}
\end{equation*}
and $\operatorname{argmin_\theta}\, \mathscr{Z}^\eps_c(\theta)$ is independent of $\eps > 0$. We also find from Figure~\ref{fig:N3_alpha_LZ}~(d) that, for any fixed $\theta \in (0,\pi/2)$, $\mathscr{Z}^\eps_c(\theta)$ seems to diverge to negative infinity as $\eps$ tends to zero, though it is not a monotonically decreasing function of $\eps$ near $\theta=0$.
Although we omit the figures, numerical computations indicate that the configuration of the three $\alpha$-point vortices at $t_c^\eps(\theta)$ is a collinear state for any $\eps>0$ and $\theta \in (0,\pi/2)$, and it is similar to \eqref{init:three-smpl} with $\theta = 0$, which is a relative equilibrium of the three PV system, see Section~\ref{subsec:exact-solutions}.

\begin{figure}[t]
\begin{center}
\includegraphics[scale=0.70]{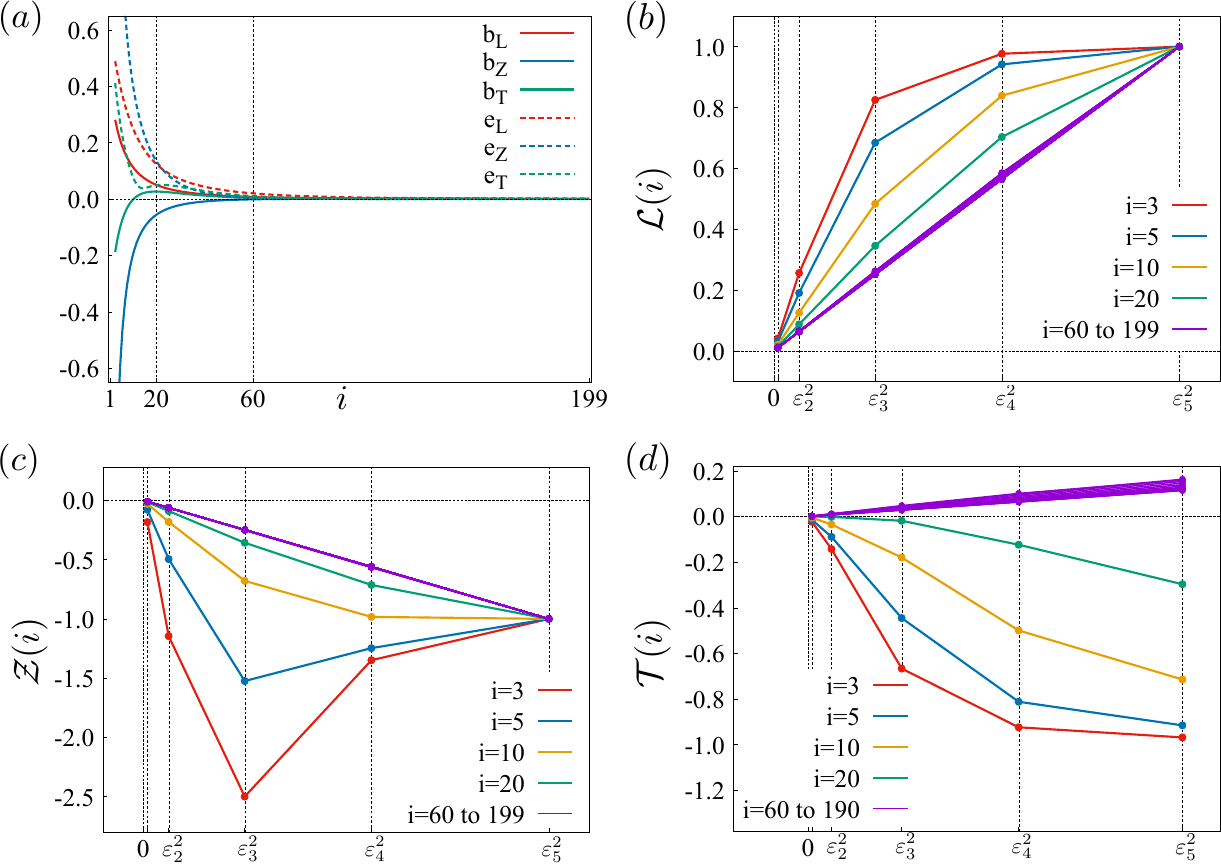}

\caption{The interpolating curves of (a) $b_L(i), b_Z(i), b_T(i)-1$ and $e_L(i), e_Z(i), e_T(i)$, $i = 1, \cdots 199$ with lines, in which the continuous curves describe $b_L, b_Z, b_T$ and the dashed ones do $e_L, e_Z, e_T$.
The plots of (b) $\mathcal{L}(i)$, (c) $\mathcal{Z}(i)$ and (d) $\mathcal{T}(i)$ with interpolating straight lines. The purple graphs in (b) and (c) are plotting data for all $i$ from $60$ to $199$ and ones in (d) are plotting skipped data for $i = j\times10$, $j=6,\cdots,19$ for visibility.}

\label{fig:N3_alpha_limit_LZT_tc}
\end{center}
\end{figure}

Next, we see the $\eps \to 0$ limits of $L^\eps_c(\theta)$, $\mathscr{Z}^\eps_c(\theta)$ and $t^\eps_c(\theta)$ for $\theta \in (0, \pi/2)$ more precisely. For $i \in \{1,\cdots,199\}$, we consider the following three sets of 5 points on $\mathbb{R}^2$,
\begin{equation}
\mathcal{L}(i) := \{ (\eps_n^2,L_n(i)) \}_{n=1}^5, \quad \mathcal{Z}(i) := \{ (\eps_n^2,Z_n(i)) \}_{n=1}^5, \quad \mathcal{T}(i) := \{ (\eps_n^2,T_n(i))\}_{n=1}^5,   \label{def:dis-points}
\end{equation}
where $\{\eps_n \}_{n=1}^5$ is given by \eqref{def:eps-n} and $L_n$, $Z_n$, $T_n$ are defined by
\begin{equation*}
L_n(i) := \frac{L^{\eps_n}_c(\theta_i)^2}{L^{\eps_5}_c(\theta_i)^2}, \qquad Z_n(i) := \frac{1/ \mathscr{Z}^{\eps_n}_c(\theta_i)}{1/|\mathscr{Z}^{\eps_5}_c(\theta_i)|} = \frac{|\mathscr{Z}^{\eps_5}_c(\theta_i)|}{\mathscr{Z}^{\eps_n}_c(\theta_i)}, \qquad T_n(i) := \frac{t^{\eps_n}_c(\theta_i)}{t_c(\theta_i)} - 1.
\end{equation*}
Here, $t_c(\theta)$ is the collapse time \eqref{def:tc-theta} of the PV system. Note that we use normalized values of $L^{\eps_n}_c(\theta)$, $1/\mathscr{Z}^{\eps_n}_c(\theta)$ and $t^{\eps_n}_c(\theta)$ divided by $L^{\eps_5}_c(\theta)$, $1/|\mathscr{Z}^{\eps_5}_c(\theta)|$ and $t_c(\theta)$, respectively.
Our purpose is to show that $\mathcal{L}(i)$, $\mathcal{Z}(i)$ and $\mathcal{T}(i)$ are on curves connected to the origin on $\R^2$ for any $i = 1,\cdots,199$,
which indicates that $L^\eps_c(\theta)$ converges to zero, $\mathscr{Z}^\eps_c(\theta)$ diverges to negative infinity and $t^\eps_c(\theta)$ converges to $t_c(\theta)$ in the $\eps \to 0$ limit.
We apply the least squares method to $\mathcal{L}(i)$, $\mathcal{Z}(i)$ and $\mathcal{T}(i)$ and try to approximate these sets by straight lines. We describe the approximate lines by
\begin{equation}
y = a_L(i) x + b_L(i), \qquad y = a_Z(i) x + b_Z(i), \qquad y = a_T(i) x + b_T(i), \label{def:ax+b}
\end{equation}
and the errors between the approximate lines and the three sets by
\begin{equation}
\begin{aligned}
& e_L(i):= \left( \sum_{n=1}^5 \left( L_n(i) - \left( a_L(i) \eps_n^2 + b_L(i) \right) \right)^2 \right)^{1/2}, \\
& e_Z(i):= \left( \sum_{n=1}^5 \left( Z_n(i) - \left( a_Z(i) \eps_n^2 + b_Z(i) \right) \right)^2 \right)^{1/2}, \\
& e_T(i):= \left( \sum_{n=1}^5 \left( T_n(i) - \left( a_T(i) \eps_n^2 + b_T(i) \right) \right)^2 \right)^{1/2},
\end{aligned}
\label{def:error}
\end{equation}
respectively. Figure~\ref{fig:N3_alpha_limit_LZT_tc}~(a) shows the graphs of $b_L(i), b_Z(i), b_T(i)$ and $e_L(i), e_Z(i), e_T(i)$ for $i = 1, \cdots, 199$.
For large $i$, $\mathcal{L}(i)$, $\mathcal{Z}(i)$ and $\mathcal{T}(i)$ are well approximated by straight lines connected to the origin since both $(b_L(i), b_Z(i),b_T(i))$ and $(e_L(i), e_Z(i),e_T(i))$ are sufficiently small.
Although the errors $e_L(i)$, $e_Z(i)$ and $e_T(i)$ for small $i$ are large, which means that $\mathcal{L}(i)$, $\mathcal{Z}(i)$ and $\mathcal{T}(i)$ are not on any straight line, these three sets are still on curves connected to the origin, see Figure~\ref{fig:N3_alpha_limit_LZT_tc}~(b), (c) and (d).
Hence, we find from the numerical computations that the desired convergences of $L^\eps_c(\theta)$, $\mathscr{Z}^\eps_c(\theta)$ and $t^\eps_c(\theta)$ hold and conclude that the enstrophy dissipation occurs by the triple collapse of point vortices. We again remark that these numerical results are consistent with the mathematical results in \cite{GS2,GS3}.

\subsection{Four vortex problem}
\label{subsec:results-four}

\begin{figure}[t]
\begin{center}

\includegraphics[scale=0.70]{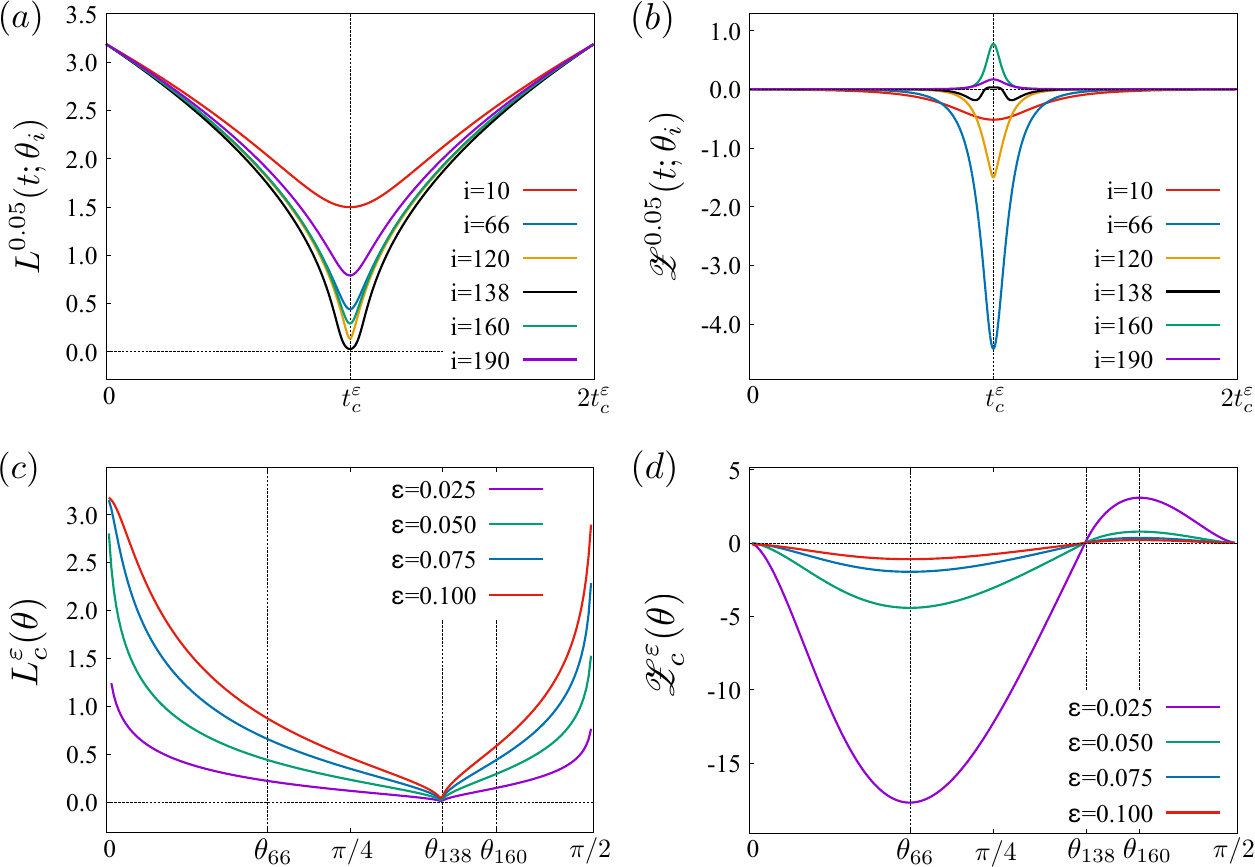}

\caption{The graphs of (a) $\{ L^{\eps_3}(t; \theta_i) \}_{i \in \mathcal{I}}$, (b) $\{ \mathscr{Z}^{\eps_3}(t;\theta_i) \}_{i \in \mathcal{I}}$ with $\mathcal{I}= \{10, 66, 120, 138, 160, 190 \}$,
(c) $\{ L^{\eps_n}_c(\theta)\}_{n=2}^5$ and (d) $\{ \mathscr{Z}^{\eps_n}_c(\theta) \}_{n=2}^5$ for the four $\alpha$PV system. Similarly to Figure~\ref{fig:N3_alpha_LZ}, the time axes are rescaled in (a) and (b), and  the graphs of (c) and (d) are interpolating curves for $i = 1, \cdots, 199$.}

\label{fig:N4_alpha_LZ}
\end{center}
\end{figure}

\begin{figure}[t]
\begin{center}
\includegraphics[scale=0.7]{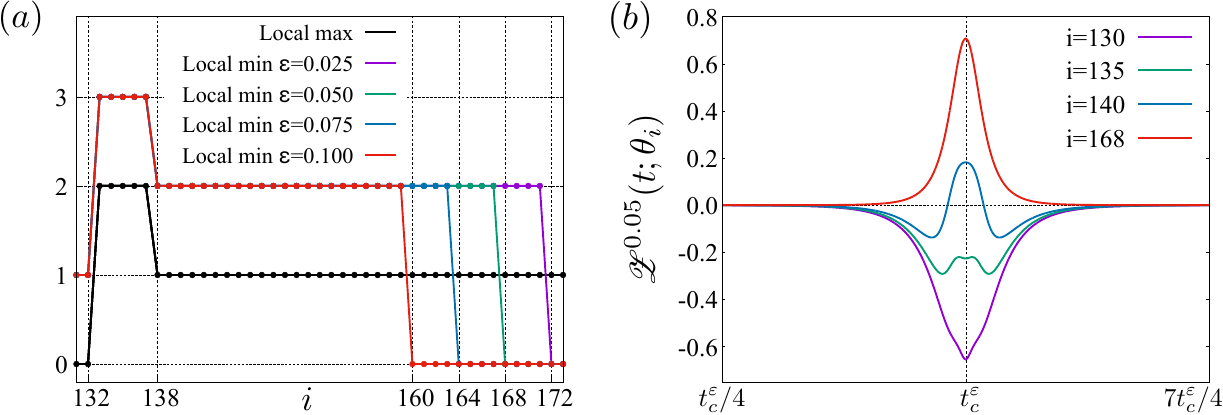}
\caption{(a) The numbers of local maximum and tlocal minimum of $\mathscr{Z}^{\eps}(t;\theta_i)$ for $i =131,\cdots, 173$. (b) The graphs of $\{ \mathscr{Z}^{\eps_3}(t;\theta_i) \}_{i \in \{130,135,140,168 \}}$.}
\label{fig:N4_alpha_extremum}

\vspace{5mm}

\includegraphics[scale=0.7]{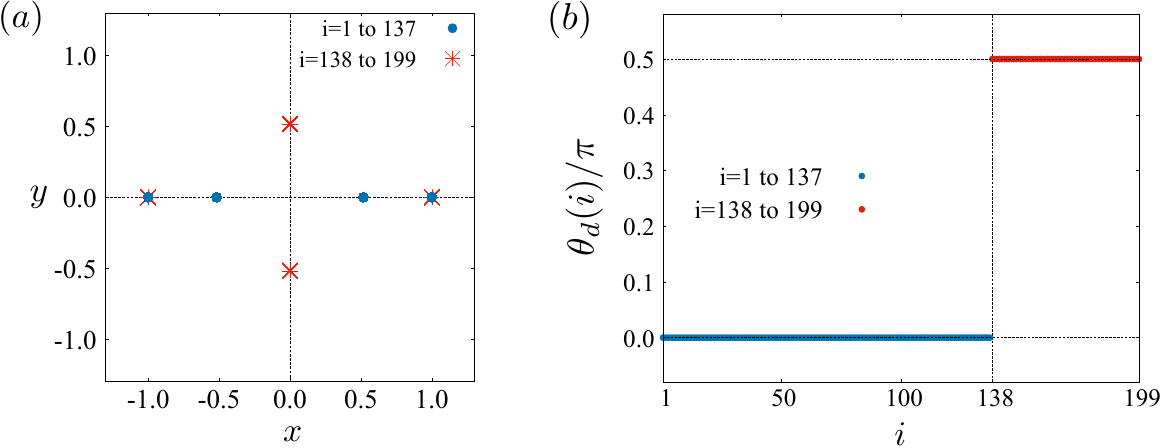}
\caption{(a) The rescaled configurations of the four $\alpha$-point vortices at $t_c^{\eps_3}(\theta_i)$. (b) The angle between the diagonals $l^{\eps_3}_{12}(t_c^{\eps_3};\theta_i)$ and $l^{\eps_3}_{34}(t_c^{\eps_3};\theta_i)$ divided by $\pi$.}
\label{fig:N4_alpha_pvc_config}

\end{center}
\end{figure}

We consider the four $\alpha$PV system with initial data \eqref{init:Novikov} without $k_5$, that is,
\begin{equation}
k_1 = \dfrac{1}{2} d_1 e^{i\theta}, \qquad k_2 = - \dfrac{1}{2} d_1 e^{i\theta}, \qquad k_3 = - \dfrac{1}{2} d_2, \qquad k_4 = \dfrac{1}{2} d_2 \label{init:four}
\end{equation}
for $\theta \in (0, \pi/2)$. For numerical computations, we set $d_2 = 2$ and $\alpha = -1$, which determine the other parameters $d_1$ and $\gamma_2$ by the relation $d_1^2 / d_2^2 = 2 - \sqrt{3}$.
As we see in Figure~\ref{fig:N4_alpha_LZ}~(a) and (c), similarly to the three vortex problem, $L^{\eps_3}(t;\theta_i)$ decreases as time evolves and, after attaining its minimum at $t_c^\eps$, it turns to increase, which holds for the other $\eps_n$ as well.
At the critical time, $L^{\eps_n}_c(\theta)$ attains its minimum at a certain $\theta_L \in (0, \pi/2)$ and it is monotonically decreasing for $\theta < \theta_L$ and increasing for $\theta > \theta_L$, which is a feature different from the three vortex problem. Numerical computations indicate $\theta_{136} < \theta_L < \theta_{138}$ for any $\eps_n$. As $\eps >0$ gets smaller, $L^\eps_c(\theta)$ seems to converge to zero: the four $\alpha$-point vortices simultaneously collapse at a finite time in the $\eps \to 0$ limit.
Regarding the enstrophy, Figure~\ref{fig:N4_alpha_LZ}~(b) and (d) show that $\mathscr{Z}^\eps(t;\theta)$ could be positive for $\theta$ larger than a certain value $\theta_Z \in (0, \pi/2)$ in contrast to the three vortex problem.
More precisely, it is suggested that $\mathscr{Z}^\eps(t;\theta)$ is a negative function of $t$ for any fixed $\theta < \theta_Z$, but $\mathscr{Z}^\eps(t;\theta)$ for $\theta > \theta_Z$ become positive around $t_c^\eps(\theta)$ and it is a positive function for sufficiently large $\theta$.
In addition, numerical computations show that, as a function of $t$, $\mathscr{Z}^\eps(t, \theta_i)$ has the one local extremum at $t_c^\eps(\theta_i)$ for $i \leq 132$ and $i \geq 173$, that is, the value $\mathscr{Z}^\eps_c(\theta_i)$ is the global minimum for $i \leq 132$ and the global maximum for $\theta \geq 173$.
For the case of $132 < i < 173$, $\mathscr{Z}^\eps(t;\theta_i)$ is in a transition process: the critical time $t_c^\eps(\theta_i)$ is not just one extremum but $\mathscr{Z}^\eps(t;\theta_i)$ has several extrema, see Figure~\ref{fig:N4_alpha_extremum}~(a) and (b).
Focusing on the critical time, $\mathscr{Z}^\eps_c(\theta)$ attains its minimum around $\theta = \theta_{66}$ and maximum around $\theta = \theta_{160}$.
The sign of $\mathscr{Z}^\eps_c(\theta)$ changes at $\theta = \theta_Z$ satisfying $\theta_{137} < \theta_Z < \theta_{138}$. Considering the $\eps \to 0$ limit, $\mathscr{Z}^\eps_c(\theta)$ seems to diverge to negative infinity for $\theta < \theta_Z$ and positive infinity $\theta > \theta_Z$, respectively. Namely, the enstrophy dissipation by collapse of the four $\alpha$-point vortices in the $\eps \to 0$ limit occurs for $\theta < \theta_Z$.
It is numerically suggested that $\theta_Z$ is the same value as $\theta_L$ which we describe by $\theta_c$. Then, the critical angle $\theta_c$ is a universal constant with respect to $\eps>0$.

We remark the configuration of the four $\alpha$-point vortices at the critical time.
Figure~\ref{fig:N4_alpha_pvc_config}~(a) shows the rescaled configurations at $t_c^{\eps_3}(\theta_i)$ for $i=1,\cdots,199$ and Figure~\ref{fig:N4_alpha_pvc_config}~(b) does the angle between $l^{\eps_3}_{12}(t_c^{\eps_3};\theta_i)$ and $l^{\eps_3}_{34}(t_c^{\eps_3};\theta_i)$ divided by $\pi$, which is denoted by $\theta_d(i)/\pi$.
For any $i = 1,\cdots,137$, the configuration at the critical time is a collinear state whose enstrophy $\mathscr{Z}^\eps_c(\theta_i)$ is negative, see Figure~\ref{fig:N4_alpha_LZ} (d). For $i = 138,\cdots,199$, the four $\alpha$-point vortices form a rhombus at $t_c^\eps(\theta_i)$ and $\mathscr{Z}^\eps_c(\theta_i)$ has a positive value, which has never been observed for the three vortex problem.
Thus, $\theta_c$ is also critical in terms of the configuration at the critical time.
It is noteworthy that the collinear and the rhombus states observed in Figure~\ref{fig:N4_alpha_pvc_config}~(a) are the same as \eqref{init:four} with $\theta = 0$ and $\theta=\pi/2$, that is, relative equilibria of the four PV system.
We have numerically obtained the same figures as Figure~\ref{fig:N4_alpha_pvc_config} for the other $\eps_n$. Thus, the enstrophy dissipation could occur by vortex collapse in the $\eps \to 0$ limit of the four $\alpha$-point vortices keeping a collinear configuration.

\begin{figure}[t]
\begin{center}
\includegraphics[scale=0.70]{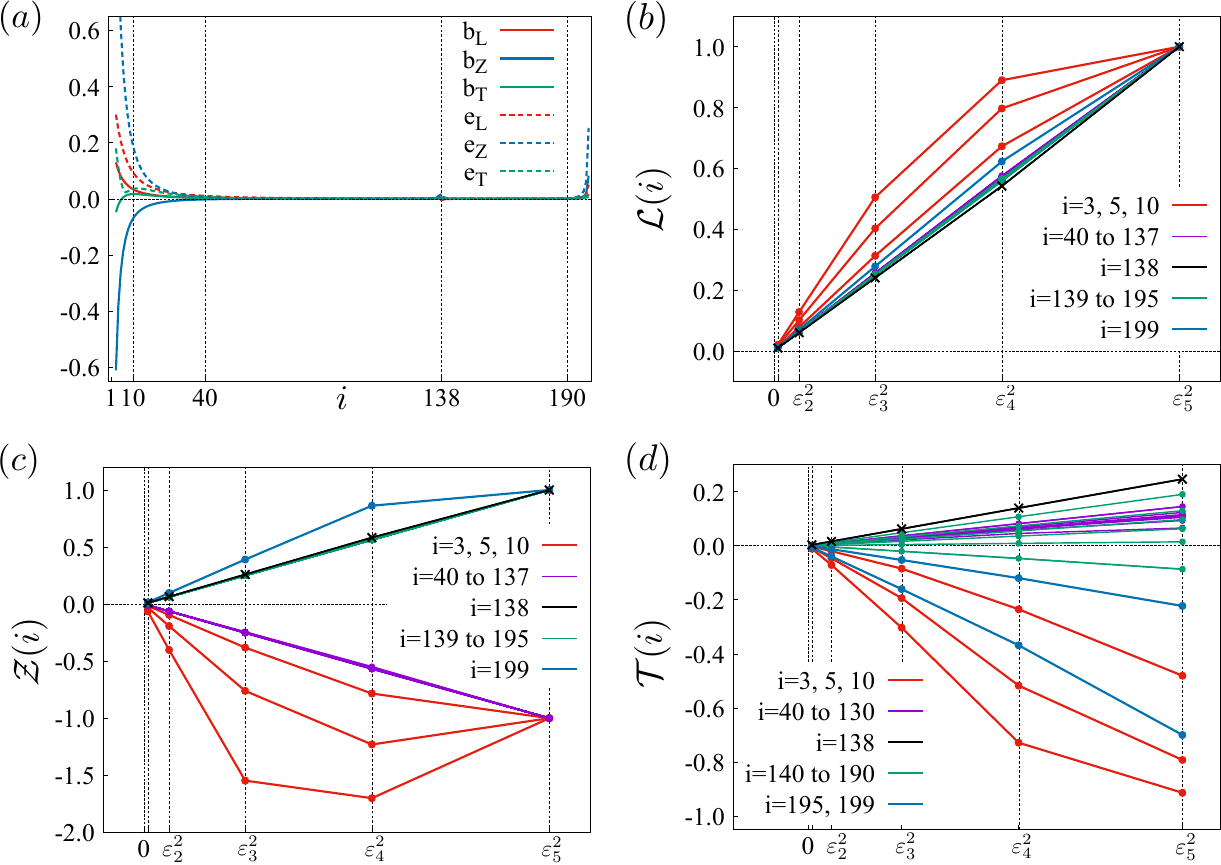}

\caption{The interpolating curves of (a) $b_L(i), b_Z(i), b_T(i)$ and $e_L(i), e_Z(i), e_T(i)$, $i = 1, \cdots 199$ with lines. The plots of (b) $\mathcal{L}(i)$, (c)$\mathcal{Z}(i)$ and (d) $\mathcal{T}(i)$ with lines. The purple and green graphs in (b) and (c) are plotting data for all $i$ in the described ranges and ones in (d) are plotting skipped data for $i = j\times10$, $j=4,\cdots,19$.}

\label{fig:N4_alpha_limit_LZT_tc}
\end{center}
\end{figure}

Next, we investigate the $\eps \to 0$ limits of $L^\eps_c(\theta)$, $\mathscr{Z}^\eps_c(\theta)$ and $t^\eps_c(\theta)$.
In the same manner as the three vortex problem, we consider the three sets $\mathcal{L}(i)$, $\mathcal{Z}(i)$ and $\mathcal{T}(i)$ in \eqref{def:dis-points} and show that $\mathcal{L}(i)$, $\mathcal{Z}(i)$ and $\mathcal{T}(i)$ are on curves connected to the origin for any $i=1,\cdots,199$. We also use the same notations about the approximate lines \eqref{def:ax+b} and the errors \eqref{def:error} based on the least squares method.
Figure~\ref{fig:N4_alpha_limit_LZT_tc} (a) shows graphs of $b_L(i), b_Z(i), b_T(i)$ and $e_L(i), e_Z(i), e_T(i)$ for $i = 1, \cdots 199$. Except for $\theta_i$ near $i =0$ and $i=199$, the three curves interpolating $\mathcal{L}(i)$, $\mathcal{Z}(i)$ and $\mathcal{T}(i)$ with lines are approximated by straight lines and the errors $b_L(i)$, $b_Z(i)$, $b_T(i)$ are sufficiently small.
Note that, for $\theta_i$ around the critical angle $\theta_c$, absolute values of $b_L$, $b_Z$, $e_L$ and $e_Z$ are slightly bigger than zero, but they are still well approximated by straight lines. As for $\theta_i$ near $i=0$ and $i=199$, the errors $e_L(i)$, $e_Z(i)$ and $e_T(i)$ are large and thus the three sets are not on any straight line. However, they are on curves connected to the origin, see Figure~\ref{fig:N4_alpha_limit_LZT_tc}~(b), (c) and (d).
Hence, numerical computations indicate that, for any $\theta \in (0, \pi/2)$, the solution to the four $\alpha$PV system with \eqref{init:four} converges to a collapsing orbit with the collapse time $t_c(\theta)$ and the enstrophy diverges to infinity: it diverges to negative infinity for $\theta < \theta_c$ and positive infinity for $\theta > \theta_c$.

\begin{figure}[t]
\begin{center}
\includegraphics[scale=0.70]{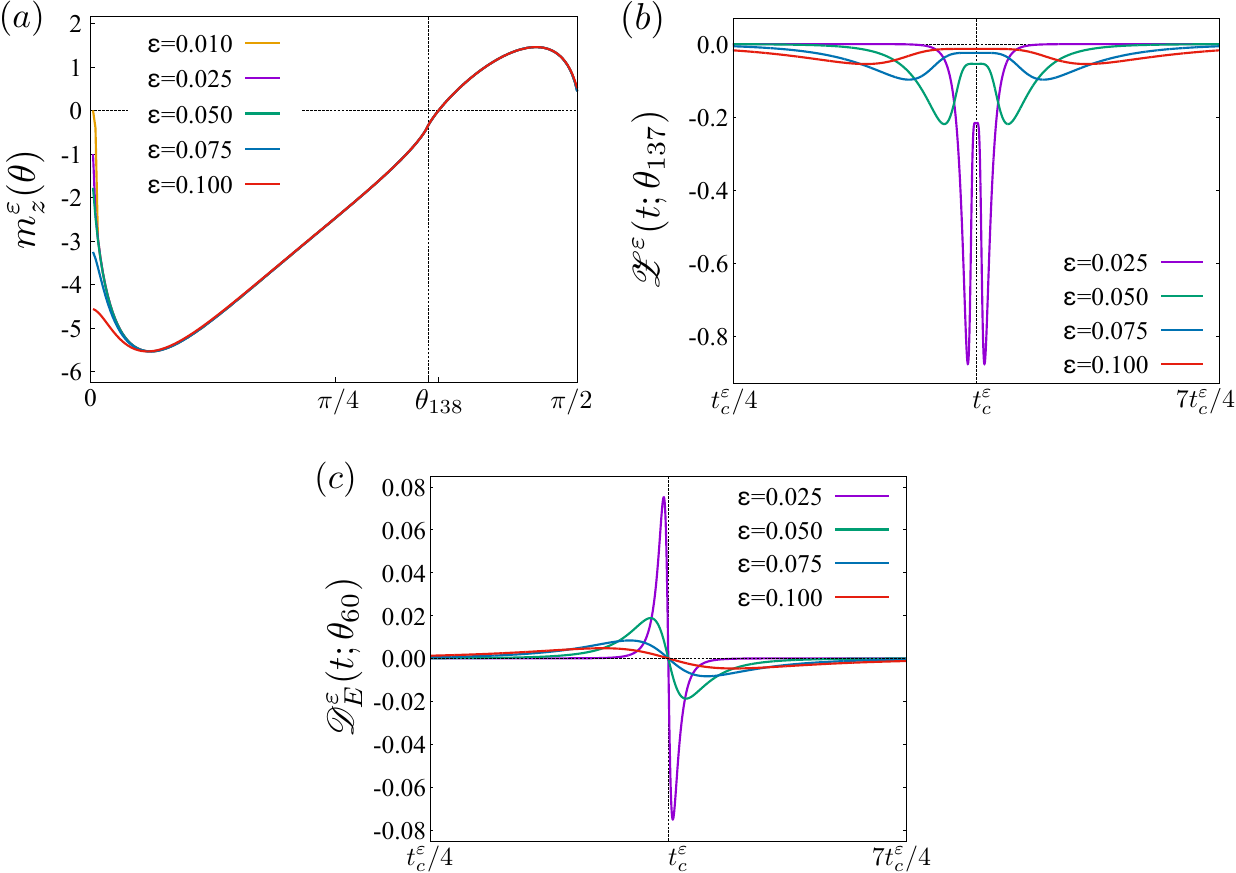}
\caption{The graphs of (a) $\{ m_z^{\eps_n}(\theta) \}_{n=1}^5$, (b) $\{ \mathscr{Z}^{\eps}(t,\theta_{137}) \}_{n=2}^5$ and (c) $\{ \mathscr{D}_E^{\eps}(t,\theta_{60}) \}_{n=2}^5$ for the four $\alpha$PV system.}
\label{fig:N4_alpha_conv_Z}
\end{center}
\end{figure}

Finally, we show that the convergences \eqref{thm:lmn} and \eqref{thm:Z-DE} in the sense of distributions hold for the four vortex problem with $\theta \in (0, \theta_c)$. The above analysis about the $\eps \to 0$ limit has already shown \eqref{thm:lmn} for $\theta \in (0, \pi/2)$.
To see that $\mathscr{Z}^\eps$ converges to the Dirac delta function, it is sufficient to show that, for any $\theta \in (0, \theta_c)$, there exists a constant $m_z(\theta) < 0$ such that
\begin{equation*}
\lim_{\eps \to 0} m_z^\eps(\theta) = m_z(\theta), \qquad m_z^\eps(\theta):=\int_{-\infty}^\infty \mathscr{Z}^\eps(t; \theta) dt,
\end{equation*}
since we have already confirmed
\begin{equation}
\lim_{\eps \to 0} \mathscr{Z}^\eps_c(\theta) =  \left\{
\begin{array}{ll}
- \infty & (\theta < \theta_c), \\
+ \infty & (\theta > \theta_c)
\end{array}
\right.  \label{lim:Z-point}
\end{equation}
and
\begin{equation}
\lim_{\eps \to 0} t_c^\eps(\theta) = t_c(\theta) \label{lim:tc}
\end{equation}
for any $\theta \in (0, \pi/2)$. As we see in Figure~\ref{fig:N4_alpha_conv_Z}~(a), for any fixed $\theta \in (0, \pi/2)$, $m_z^\eps(\theta)$ converges to a certain value as $\eps$ tends to zero. Thus, defining the function $m_z(\theta)$ by the limit, we obtain the convergence to the Dirac delta function.
Note that $\mathscr{Z}^\eps(t,\theta)$ has several local minima for $\theta \in (\theta_{132}, \theta_c)$, see Figure~\ref{fig:N4_alpha_extremum}~(a), and the convergence to the Dirac delta function is not obvious.
However, as we see in Figure~\ref{fig:N4_alpha_conv_Z}~(b), the times when $\mathscr{Z}^\eps(t,\theta_{137})$ attains its local minima get close to $t_c^\eps(\theta_{137})$ as $\eps$ tends to zero, that is, it converges to the collapse time $t_c(\theta_{137})$ owing to \eqref{lim:tc}. Since numerical computations show that the same result holds for $\theta_i$, $i=133,\cdots,136$,
 we find from \eqref{lim:Z-point} that the desired convergence holds for any $\theta \in (\theta_{132}, \theta_c)$.
As for the convergence of the energy dissipation rate $\mathscr{D}_E^\eps$, it is enough to show that $\mathscr{D}_E^\eps(t-t_c^\eps)$ is a odd and integrable function on $\R$, see the proof of Theorem~6 in \cite{GS3}. As an example, Figure~\ref{fig:N4_alpha_conv_Z}~(c) shows the graph of $\mathscr{D}_E^\eps(t-t_c^\eps; \theta_{60})$ and we find that it is odd and rapidly decreasing as $t$ gets further away from $t_c^\eps(\theta_{60})$ for any $\eps_n$.
Although we omit the figures, numerical computations show that $\mathscr{D}_E^\eps(t-t_c^\eps;\theta_i)$ is odd and integrable for any $i=1, \cdots, 199$. Thus, we conclude that, for any $\theta \in (0, \theta_c)$, $\mathscr{Z}^\eps(\cdot; \theta)$ converges to the Dirac delta function with the mass $m_z(\theta) < 0$ and the support $t= t_c(\theta)$, and $\mathscr{D}_E^\eps(\cdot;\theta)$ converges to zero in the sense of distributions.

\subsection{Five vortex problem}
\label{subsec:results-five}

\begin{figure}[t]
\begin{center}
\includegraphics[scale=0.70]{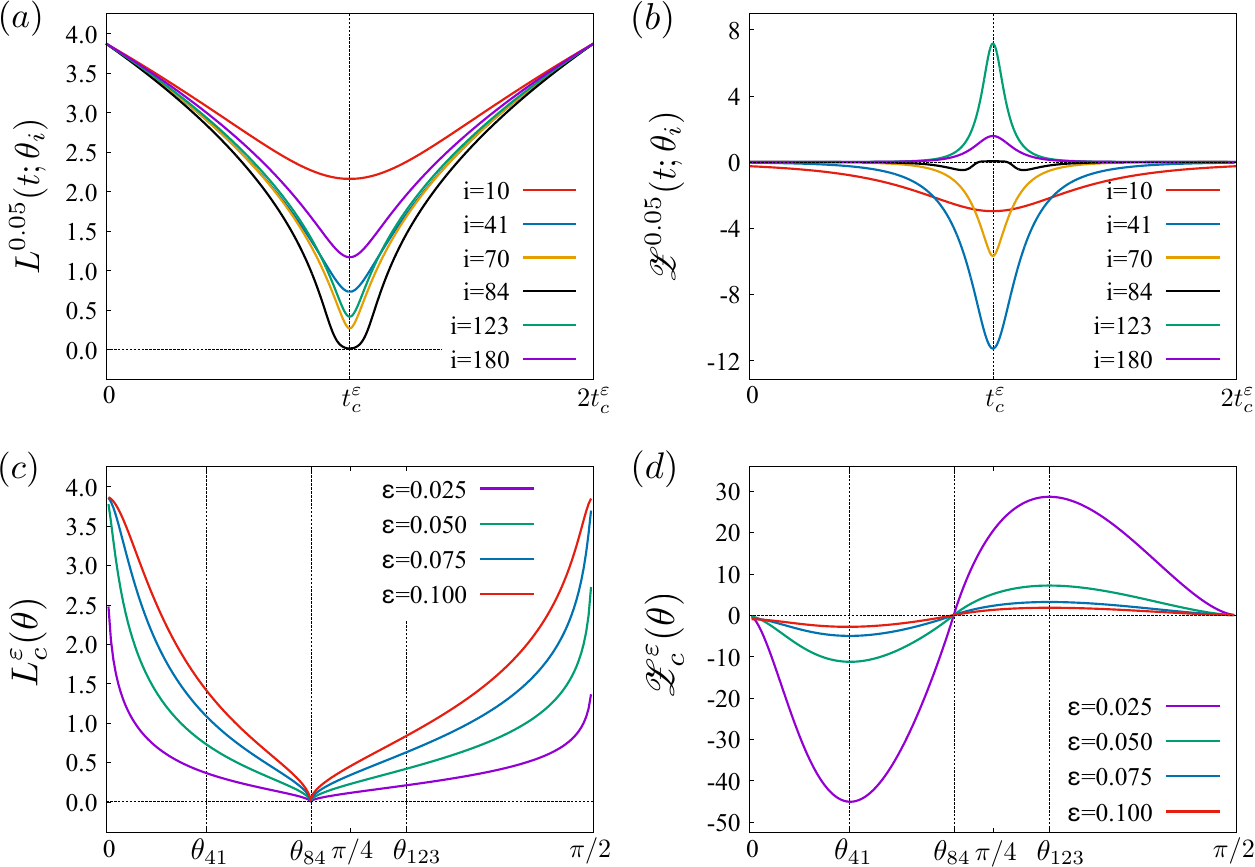}

\caption{The graphs of (a) $\{ L^{\eps_3}(t; \theta_i) \}_{i \in \mathcal{I}}$, (b) $\{ \mathscr{Z}^{\eps_3}(t;\theta_i) \}_{i \in \mathcal{I}}$ with $\mathcal{I}= \{10, 41, 70, 84, 123, 180 \}$,
(c) $\{ L^{\eps_n}_c(\theta)\}_{n=2}^5$ and (d) $\{ \mathscr{Z}^{\eps_n}_c(\theta) \}_{n=2}^5$ for the five $\alpha$PV system. Similarly to Figure~\ref{fig:N3_alpha_LZ} and \ref{fig:N4_alpha_LZ}, the time axes are rescaled in (a) and (b), and  the graphs of (c) and (d) are interpolating curves for $i = 1, \cdots, 199$.}

\label{fig:N5_alpha_LZ}
\end{center}
\end{figure}

We consider the five vortex problem with initial data \eqref{init:Novikov} for $\theta \in (0,\pi/2)$. In the following numerical computations, we use the parameters $d_2 = 2$, $\alpha = -1$ and $\beta = 1/2$. Then, $d_1$ and $\gamma$ are determined by the relations $I = M = \Gamma_H=0$.
As we see in Figure~\ref{fig:N5_alpha_LZ}~(a) and (b), the functions $L^{\eps}(t;\theta_i)$ and $\mathscr{Z}^\eps(t;\theta_i)$ behave almost the same way as the four vortex problem and numerical computations for the other $\eps_n$ show the same features. As for the values of $L^{\eps}(t;\theta)$ and $\mathscr{Z}^\eps(t;\theta)$ at the critical time,
Figure~\ref{fig:N5_alpha_LZ}~(c) and (d) show that there exists a critical angle $\theta_c$ in $(\theta_{83},\theta_{84})$, which is universal with respect to $\eps>0$, such that $L^\eps_c(\theta)$ attains its minimum at $\theta = \theta_c$ and $\mathscr{Z}^\eps_c(\theta)$ changes its sign before and after $\theta_c$.
In addition, $\mathscr{Z}^\eps_c(\theta)$ has the global minimum around $\theta_{41}$ and the global maximum around $\theta_{123}$ independent of $\eps_n$.
The above features are similar to the four vortex problem but the critical angle $\theta_c$ is smaller than one in the four vortex problem, that is, the enstrophy dissipation is less likely to occur in the five vortex problem.

\begin{figure}[t]
\begin{center}
\includegraphics[scale=0.7]{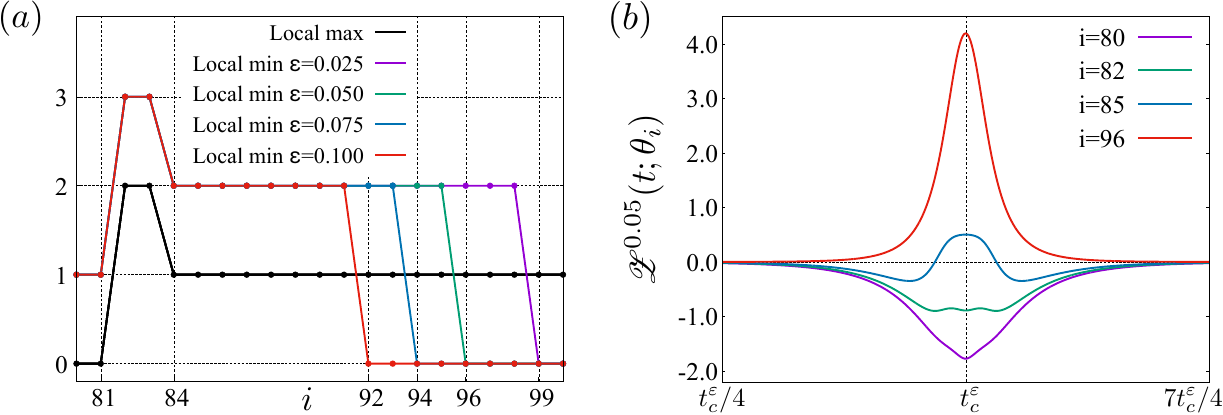}
\caption{(a) The numbers of local maximum and local minimum of $\mathscr{Z}^{\eps}(t;\theta_i)$ for $i =80,\cdots, 100$. (b) The graphs of $\{ \mathscr{Z}^{\eps_3}(t;\theta_i) \}_{i \in \{80,82,85,96 \}}$.}
\label{fig:N5_alpha_extremum}

\vspace{6mm}

\includegraphics[scale=0.7]{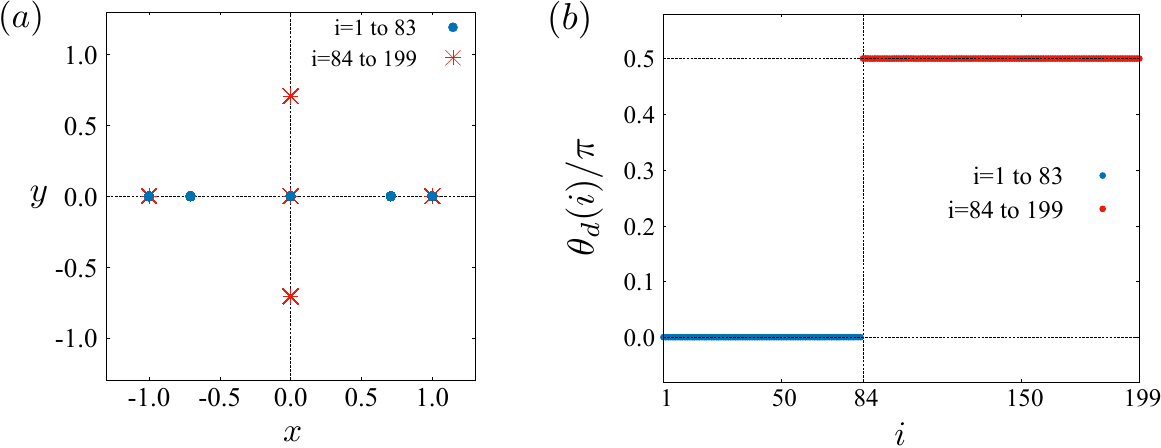}
\caption{(a) The rescaled configurations of the five $\alpha$-point vortices at $t_c^{\eps_3}(\theta_i)$. (b) The angle between the diagonals $l^{\eps_3}_{12}(t_c^{\eps_3};\theta_i)$ and $l^{\eps_3}_{34}(t_c^{\eps_3};\theta_i)$ divided by $\pi$.}
\label{fig:N5_alpha_pvc_config}

\end{center}
\end{figure}

As for the behavior of $\mathscr{Z}^\eps(t, \theta_i)$ for $\theta_i$ around $\theta_c$, similarly to the four vortex problem, the number of its extremum is not single and there is a transition process, see Figure~\ref{fig:N5_alpha_extremum} for the detail.
The configuration of the five $\alpha$-point vortices at the critical time is also changes before and after $\theta_c$: configurations for $\theta < \theta_c$ are collinear states and those for $\theta > \theta_c$ are rhombuses, see Figure~\ref{fig:N5_alpha_pvc_config}.
Note that the above collinear and rhombus states are similar to relative equilibria \eqref{init:Novikov} with $\theta = 0$ and $\theta=\pi/2$ in the five PV system.
Since we have obtained the same result as Figure~\ref{fig:N5_alpha_pvc_config} for the other $\eps_n$,
the enstrophy dissipation is caused by the collapse in the $\eps \to 0$ limit of the five $\alpha$-point vortices keeping a collinear configuration as well as the four vortex problem.

\begin{figure}[t]
\begin{center}
\includegraphics[scale=0.70]{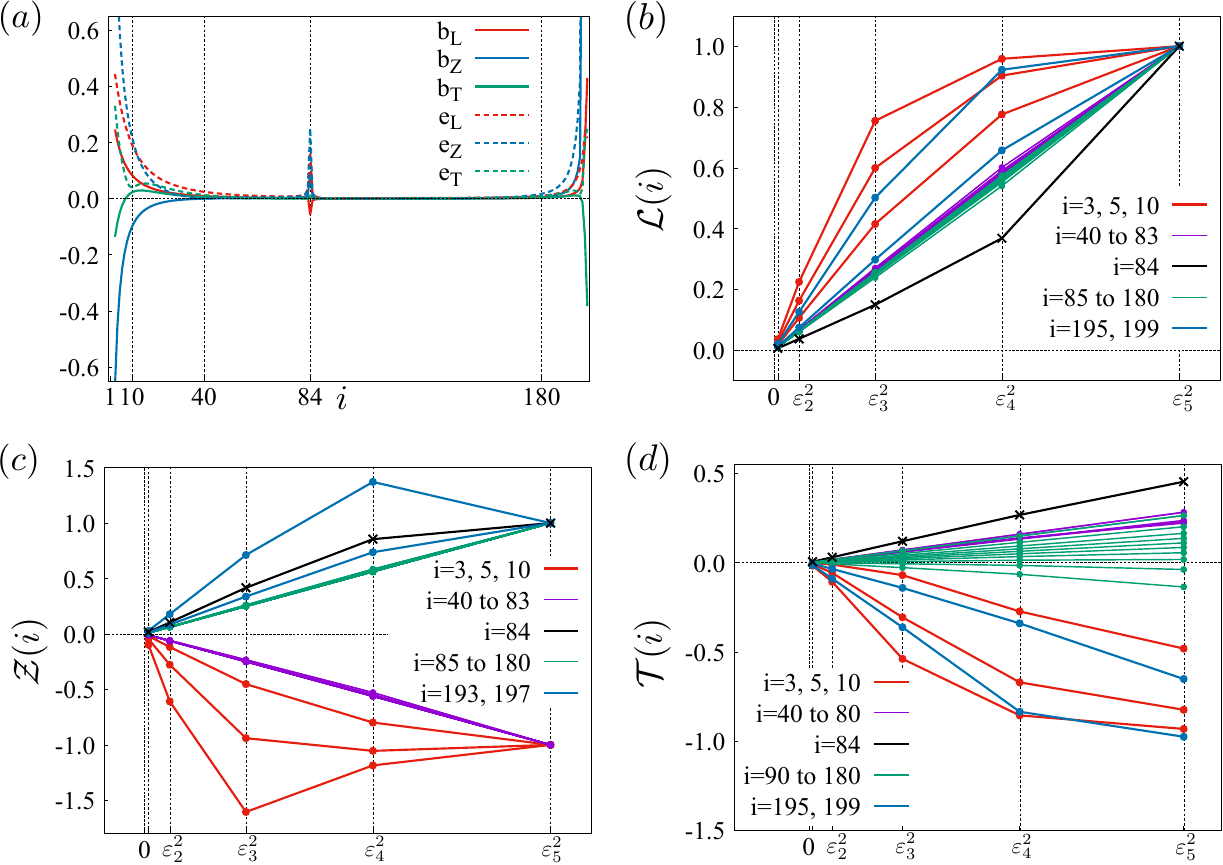}

\caption{The interpolating curves of (a) $b_L(i), b_Z(i), b_T(i)$ and $e_L(i), e_Z(i), e_T(i)$, $i = 1, \cdots 199$ with lines. The plots of (b) $\mathcal{L}(i)$, (c)$\mathcal{Z}(i)$ and (d) $\mathcal{T}(i)$ with lines. The purple and green graphs in (b) and (c) are plotting data for all $i$ in the described range and ones in (d) are plotting skipped data for $i = j\times10$, $j=4,\cdots,18$.}

\label{fig:N5_alpha_limit_LZT_tc}
\end{center}
\end{figure}

We investigate the $\eps \to 0$ limits of $L^\eps_c(\theta)$, $\mathscr{Z}^\eps_c(\theta)$ and $t^\eps_c(\theta)$ by considering
$\mathcal{L}(i)$, $\mathcal{Z}(i)$ and $\mathcal{T}(i)$ in \eqref{def:dis-points} and using the notations \eqref{def:ax+b} and \eqref{def:error} based on the least squares method.
As we see in Figure~\ref{fig:N5_alpha_limit_LZT_tc} (a), except for $\theta_i$ near $i =0$, $i=84$ and $i=199$, the three curves interpolating $\mathcal{L}(i)$, $\mathcal{Z}(i)$ and $\mathcal{T}(i)$ are approximated by straight lines and $b_L(i)$, $b_Z(i)$, $b_T(i)$ are sufficiently close to zero.
Although the errors $e_L(i)$, $e_Z(i)$ and $e_T(i)$ near $i =0$, $i =84$ and $i=199$ are large and their data are not on any straight line, they are on curves connected to the origin, see Figure~\ref{fig:N5_alpha_limit_LZT_tc}~(b), (c) and (d).
Thus, we conclude that, for any $\theta \in (0, \pi/2)$, the solution to the five $\alpha$PV system with \eqref{init:Novikov} converges to a collapsing orbit with the collapse time $t_c(\theta)$, which it equivalent to \eqref{thm:lmn}, and the enstrophy diverges to negative infinity for $\theta < \theta_c$ and positive infinity for $\theta > \theta_c$.

\begin{figure}[t]
\begin{center}
\includegraphics[scale=0.70]{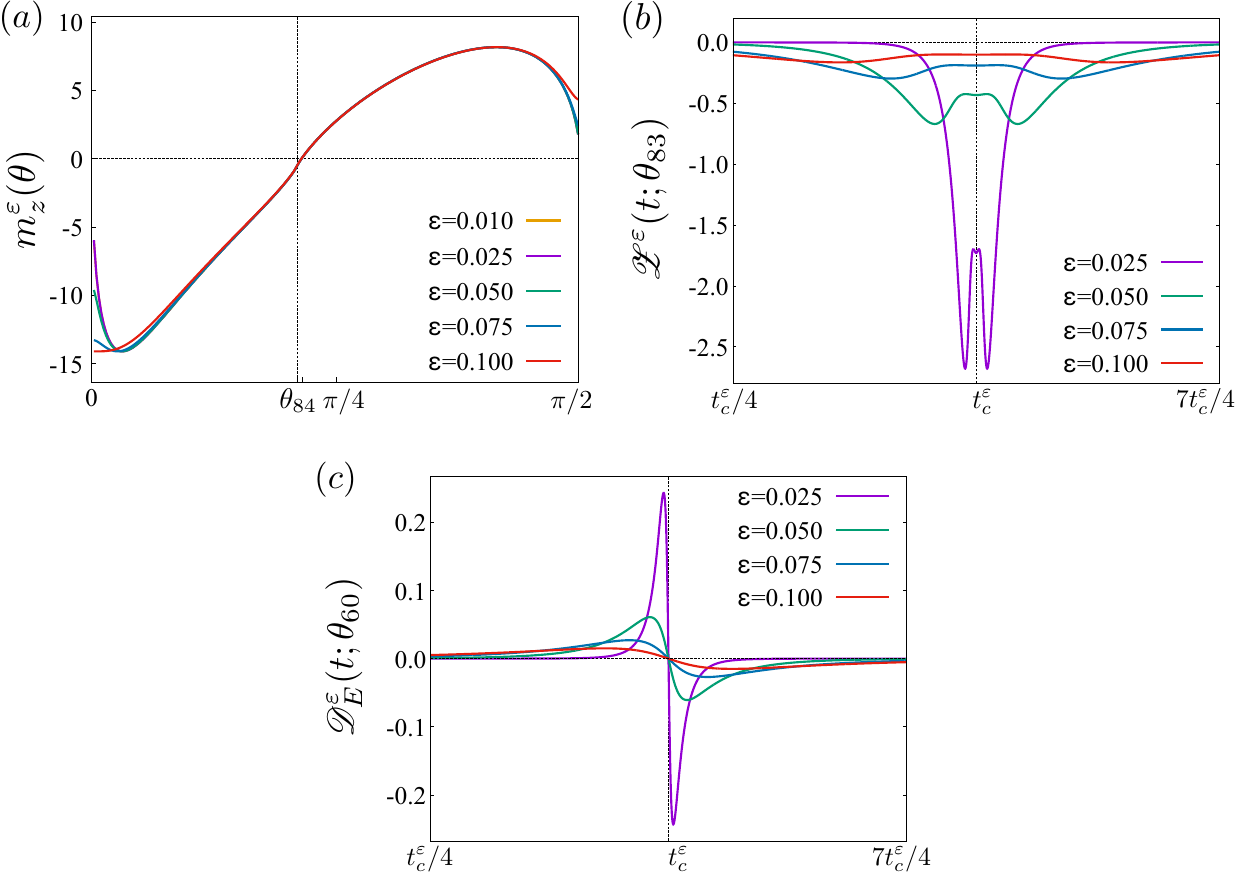}
\caption{The graphs of (a) $\{ m_z^{\eps_n}(\theta) \}_{n=1}^5$, (b) $\{ \mathscr{Z}^{\eps}(t,\theta_{83}) \}_{n=2}^5$ and (c) $\{ \mathscr{D}_E^{\eps}(t,\theta_{60}) \}_{n=2}^5$ for the five $\alpha$PV system.}
\label{fig:N5_alpha_conv_Z}
\end{center}
\end{figure}

As for the convergence \eqref{thm:Z-DE}, Figure~\ref{fig:N5_alpha_conv_Z}~(a) shows that, for any fixed $\theta \in (0, \pi/2)$, $m_z^\eps(\theta)$ converges to a constant in the $\eps \to 0$ limit. Thus, similarly to the four vortex problem, we define the function $m_z(\theta)$ by the limit values and then find
\begin{equation*}
\lim_{\eps \to 0} m_z^\eps(\theta) = m_z(\theta)
\end{equation*}
for $\theta \in (0, \pi/2)$ and, especially, $m_z(\theta) < 0$ for $\theta \in (0,\theta_c)$. Since \eqref{lim:Z-point} follows from the analysis about the $\eps \to 0$ limit, we obtain the convergence \eqref{thm:Z-DE}.
Although Figure~\ref{fig:N5_alpha_extremum}~(a) shows that $\mathscr{Z}^\eps(t,\theta)$ has several local minima for any $\theta \in (\theta_{81}, \theta_c)$, we find from Figure~\ref{fig:N5_alpha_conv_Z}~(b) that the convergence of $\mathscr{Z}^\eps(t,\theta)$ to the Dirac delta function still holds in the same manner as the four vortex problem.
As for $\mathscr{D}_E^\eps$, Figure~\ref{fig:N5_alpha_conv_Z}~(c) and other numerical computations indicate that $\mathscr{D}_E^{\eps}(t-t_c^\eps; \theta_{60})$ is a odd and integrable function for any $\eps > 0$ and $\mathscr{D}_E^{\eps}(t-t_c^\eps; \theta_i)$ is as well for $i=1, \cdots, 199$.
Thus, we conclude that, for any $\theta \in (0, \theta_c)$, $\mathscr{Z}^\eps(\cdot; \theta)$ converges to the Dirac delta function with the negative mass $m_z(\theta) < 0$ and the support $t= t_c(\theta)$, and $\mathscr{D}_E^\eps(\cdot;\theta)$ converges to zero in the sense of distributions.

\section{Concluding remarks}
\label{sec:concluding}

We have numerically investigated the dynamics of point-vortex solutions to the 2D filtered-Euler equations with the initial data for which the solution to the PV system leads to self-similar collapse in a finite time. In particular, we have considered the three, four and five vortex problems for which explicit formulae of self-similar collapsing solutions have been obtained. In the three vortex problem, preceding results have already proven that the solution to the 2D filtered-Euler equations converges to a self-similar collapsing orbit of the three PV system and dissipates the enstrophy by the collapse in the $\eps \to 0$ limit. In this paper, we have treated the Euler-$\alpha$ equations for numerical computations and shown the detailed processes of the triple collapse and the induced enstrophy dissipation. As the main results, we have shown that the enstrophy dissipation by collapse of point vortices could occur for the four and five vortex problems in the $\eps \to 0$ limit of the 2D Euler-$\alpha$ equations. That is to say, the anomalous enstrophy dissipation by vortex collapse is not specific to three vortices and it could be universal mechanism for multiple vortices in 2D inviscid flows.

We make some remarks and mention future directions. Our numerical computations have shown that, for some initial data leading to a self-similar collapse in the PV system, the corresponding filtered-point vortices converge to a collapsing orbit, but the mass of the enstrophy variation is not negative. We have not given any physical interpretation to this phenomenon. Although, we have numerically suggested that the configuration of filtered-point vortices at the critical time is essentially related to the sign of the mass, further numerical or mathematical investigation is required. It is also necessary to investigate whether or not the same result holds for other filtered models such as the vortex blob regularization and the exponential filter and how differences among these models appear. It is a challenging attempt to find the enstrophy dissipating solutions for $N$ vortex problems with $N \geq 6$, but computing the multiple FPV system accurately is a difficult problem due to numerical errors.

\subsection*{Acknowledgements}
The author would like to thank Professor Takashi Sakajo for productive discussions and valuable comments. This work was supported by JSPS KAKENHI Grant Number JP21K13820, JP23K20808 and JP24K16960.


\end{document}